\documentclass[preprint,prd,nofootinbib,tightenlines,amssymb]{revtex4}

\usepackage[dvipdfmx]{graphicx}
\usepackage{bm}
\usepackage{amsmath}
\usepackage{hhline}
\usepackage{color}
\usepackage{xcolor}
\usepackage{slashed}
\usepackage{cancel}
\usepackage{relsize}
\usepackage[T1]{fontenc}
\usepackage{upquote}
\usepackage{changes} 

\newcommand{\Eq}{&=&}

\newcommand{\white}[1]{{\color[rgb]{1,1,1} #1}}
\newcommand{\black}[1]{{\color[rgb]{0,0,0} #1}}
\newcommand{\bs}[1]{{\boldsymbol{#1}}}
\newcommand{\hs}[1]{{\hspace{#1}}}
\newcommand{\vs}[1]{{\vspace{#1}}}
\newcommand{\tf}[1]{{\textsf{#1}^{}}}
\newcommand{\tx}[1]{{\text{#1}^{}}}
\newcommand{\nn}{\nonumber\\}

\newcommand{\sx}[2]{{\scalebox{#1}{#2}}}
\newcommand{\dd}{{\mathrm{d}}}

\newcommand{\CapiN}{C_{a\pi \hs{-0.02cm} N}}
\newcommand{\CaND}{C_{aN \hs{-0.03cm} \Delta}}
\newcommand{\Epi}{E_\pi}
\newcommand{\vkpi}{|\bs{k}^{}_\pi|}
\newcommand{\vka}{|\bs{k}^{}_a|}

\newcommand{\cQL}{{\cal Q}^{}_L}
\newcommand{\cQR}{{\cal Q}^{}_R}

\newcommand{\Hu}{H^{}_{\hs{-0.03cm}u}}
\newcommand{\Hd}{H^{}_{\hs{-0.03cm}d}}

\newcommand{\qPQ}{q^{}_\tf{PQ}}
\newcommand{\QL}{Q^{}_{\hs{-0.03cm}L}}
\newcommand{\UR}{U^{}_{\hs{-0.04cm}R}}
\newcommand{\DR}{D^{}_{\hs{-0.04cm}R}}

\newcommand{\LL}{L^{}_{\hs{-0.03cm}L}}
\newcommand{\ER}{E^{}_{\hs{-0.03cm}R}}

\newcommand{\Xq}{{\cal X}_q}
\newcommand{\Qa}{{\cal Q}^{}_a}

\newcommand{\qL}{q^{}_L}
\newcommand{\qR}{q^{}_R}

\newcommand{\Mq}{{\cal M}_q}
\newcommand{\Ma}{{\cal M}_a}

\newcommand{\Bv}{{\cal B}^{}_v}
\newcommand{\Amu}{{\cal A}^{}_\mu}

\newcommand{\Dmu}{{\cal D}^{}_{\hs{-0.05cm}\mu}}
\newcommand{\Vmu}{{\cal V}^{}_{\hs{-0.02cm}\mu}}

\newcommand{\Sv}{S^\mu_v}

\newcommand{\SUL}{\tx{SU}(3)^{}_{\hs{-0.03cm}L}}
\newcommand{\SUR}{\tx{SU}(3)^{}_{\hs{-0.03cm}R}}

\newcommand{\UL}{{\cal U}^{\white{\dagger}}_L}
\newcommand{\URd}{{\cal U}_R^\dagger}
\newcommand{\UH}{{\cal U}^{\white{\dagger}}_H}
\newcommand{\UHd}{{\cal U}_H^\dagger}

\newcommand{\Tmuu}{{\cal T}^\mu_v}
\newcommand{\Tmud}{{\cal T}^{}_{v\mu}}

\newcommand{\OL}[1]{{\overline{#1}}}

\begin{document}
\baselineskip=14.5pt \parskip=2.5pt

\vspace*{3em}

\preprint{KIAS-P22082}

\title{\large Supernova Axion Emissivity with $\Delta(1232)$ Resonance\\ in Heavy Baryon Chiral Perturbation Theory}

\author{
Shu$\,$-Yu\,\,Ho\footnote[1]{phyhunter@kias.re.kr},
Jongkuk\,\,Kim\footnote[2]{jkkim@kias.re.kr},
Pyungwon\,\,Ko\footnote[3]{pko@kias.re.kr},
Jae${}^{}{}^{}{}^{}$-hyeon\,\,Park\footnote[4]{jhpark@kias.re.kr}
}

\affiliation{
Korea Institute for Advanced Study (KIAS), Seoul 02455, Republic of Korea
\vspace{3ex}}

\begin{abstract}
In this paper, we evaluate the energy loss rate of supernovae induced by the axion emission process $\pi^- + p \to n + a$ with the $\Delta(1232)$ resonance in the heavy baryon chiral perturbation theory for the first time.\,\,Given the axion-nucleon-$\Delta$ interactions, we include the previously ignored $\Delta$-mediated graphs to the $\pi^- + p \to n + a$ process.\,\,In particular, the $\Delta^0$-mediated diagram can give a resonance contribution to the supernova axion emission rate when the center-of-mass energy of the pion and proton approaches the $\Delta(1232)$ mass.\,\,With these new contributions, we find that for the typical supernova temperatures, compared with the earlier work with the axion-nucleon (and axion-pion-nucleon contact) interactions, the supernova axion emissivity can be enhanced by a factor of $\sim$\,4\,(2) in the Kim-Shifman-Vainshtein-Zakharov model and up to a factor of $\sim$\,5\,(2) in the Dine-Fischler-Srednicki-Zhitnitsky model with small $\tan\beta$ values.\,\,Remarkably, we notice that the 
$\Delta(1232)$ resonance gives a destructive contribution to the supernova axion emission 
rate at high supernova temperatures, which is a nontrivial result in this study.
\end{abstract}

\maketitle

\section{Introduction}\label{sec:1}

The QCD axion, which is a pseudo${}^{}{}^{}$-Nambu${}^{}$-Goldstone boson associated with a spontaneous breakdown of the Peccei-Quinn (PQ) global axial symmetry~\cite{Weinberg:1977ma,Wilczek:1977pj}, is so far the most promising solution to the QCD strong $C\hs{-0.03cm}P$ problem~\cite{Cheng:1987gp}.\,\,Through the PQ mechanism~\cite{Peccei:1977hh,Peccei:1977ur}, the QCD axion starts to roll down and oscillate on its potential when the Hubble parameter falls below the mass of the QCD axion and eventually settles down at a $C\hs{-0.03cm}P$-conserving minimum, solving the strong $C\hs{-0.03cm}P$ problem dynamically.\,\,In addition, it has been shown that such a coherent oscillation of the axion field behaves as cold dark matter in the present universe~\cite{Abbott:1982af,Preskill:1982cy,Dine:1982ah}.\,\,On the other hand, it has been studied that such cold axion particles can also form a Bose${}^{}{}^{}$-Einstein condensation through their self-gravitational interactions~\cite{Sikivie:2009qn}.\,\,For recent reviews of axions, one can see Refs.\,\cite{Kim:2008hd,DiLuzio:2020wdo,Choi:2020rgn}.

The QCD axion can interact with the standard model (SM) particles such as electrons and nucleons with coupling strength as well as its mass inversely proportional to the so-called axion decay constant.\,\,This axion decay constant is related to the PQ symmetry breaking scale which is typically far above the scale of the electroweak (EW) phase transition.\,\,Thus, the QCD axion feebly couples to the SM fields due to the large decay constant.\,\,However, although the coupling strength of light axions to the matter is in the weak regime, the astrophysical observations can still place    severe constraints on these axion couplings~\cite{Raffelt:2006cw,DiLuzio:2021ysg}.\,\,This is because the axions can be copiously produced from some hot and dense celestial bodies such as supernovae, neutron stars, and white dwarfs, which in turn changes their evolution.\,\,For instance, a core-collapse supernova (SN), e.g., SN1987A, can emit axions in addition to the neutrino emission as an extra cooling process of the associated neutron star.\,\,As a result, the axion emissivity from a SN core would suppress the neutrino flux and impose stringent bounds on the axion couplings to the nucleon~\cite{Turner:1987by,Raffelt:1987yt}.

There are two hadronic processes that can generate axions inside SNe, the nucleon-nucleon bremsstrahlung process $N + N \to N + N + a\,(N = n,p)$~\cite{Brinkmann:1988vi,Iwamoto:1992jp,Carenza:2019pxu} and the pion-induced Compton like process $\pi^- + p \to n + a$~\cite{Turner:1991ax,Raffelt:1993ix,Keil:1996ju}, where $a$ is the QCD axion.\,\,The former process has been thought of as the dominant axion production in a SN core for a period, and the latter one has been ignored because of the underestimation of the pion abundance inside SNe.\,\,However, with a better description of the nuclear interaction beyond the one-pion exchange graph~\cite{Turner:1987by}, the later studies have reduced the reaction rate of the nucleon-nucleon bremsstrahlung process by orders of magnitude~\cite{Raffelt:1991pw,Hannestad:1997gc,Raffelt:1996di}.\,\,On the other hand, recent  analyses have shown that pion number yields and reactions involving pions can be enhanced inside SNe due to pion-nucleon interactions~\cite{Fore:2019wib} and medium effects~\cite{Carenza:2020cis,Fischer:2021jfm}, respectively.\,\,In the case where the pions are non-negligible in SNe, it has been demonstrated that the pion-induced Compton like process can dominate over the nucleon-nucleon bremsstrahlung to be the main source of the axion emission inside SNe. 

The axion emission rate of the pion-induced Compton like process in SNe with the medium effect was first estimated in Ref.\,\cite{Carenza:2020cis}.\,\,However, they only considered nucleon-mediated diagrams $\pi^- + p \to N^\ast \to n + a$ and somehow ignored the axion-pion-nucleon contact diagram in their calculation.\,\,It is important to keep the axion-pion-nucleon contact interaction even at zero temperature, since it is allowed by spontaneously broken chiral symmetry and the associated axial current.\,\,This missing axion emission diagram has been included in Ref.\,\cite{Choi:2021ign}, indicating that the SN axion emission rate from $\pi^- + p \to n + a$ can be enhanced by a factor of at least 2 due to the axion-pion-nucleon contact interaction.\footnote{They have ignored the background matter effect in their calculation for simplicity and left it as future work.}\,\,Meanwhile, it was pointed out by a recent paper~\cite{Vonk:2022tho} that the decuplet baryon-mediated diagram $a + N \to \Delta^{\ast} \to \pi + N$  may be potentially crucial to the pion axioproduction $a + N \to \pi + N$, which was not realized before.\,\,

In this work, we point out that the $\Delta (1232)$ resonance can make significant contributions to the SN axion emission rate, which is nothing but the reversed process, $\pi^- + p \rightarrow \Delta^\ast \rightarrow n + a$, of the pion axioproduction considered in Ref.\,\cite{Vonk:2022tho}.\,\,The reason for it is straightforward.\,\,Firstly, for the typical SN temperatures, $T \sim (30\,\,\tx{to}\,\,40)\,\tx{MeV}$, the pion momentum is $\vkpi \simeq \sqrt{3{}^{}m^{}_\pi T}  \simeq m^{}_\pi$.\,\,Hence, the pion kinetic energy inside SNe is about $E^{}_\pi = \sqrt{\vkpi^2 + m_\pi^2} \sim 200\,\tx{MeV}$.\,\,In such a case, the invariant mass of the initial $\pi^-p$ system is somewhere in the middle of $\Delta(1232)$ and nucleon masses.\,\,Therefore, we cannot turn a blind eye to the $\Delta(1232)$ contributions for the SN axion emissivity.\,\,In this work, we then include $\Delta(1232)$ baryon in the intermediate state with the virtual $N$, $\pi^- + p \to (N^\ast, \Delta^\ast) \to n + a$, and the axion-pion-nucleon contact graph to the SN axion emission rate of the pion-induced Compton like channel.\,\,Depending on the couplings and signs of various terms, the $\Delta(1232)$ contributions could interfere with the virtual $N$ and axion-pion-nucleon contact term contributions either constructively or destructively.\,\,Correspondingly, the resulting constraints on the axion coupling (or equivalently, decay constant) could be either stronger or weaker.\,\,It is crucial to evaluate the amplitude for the underlying process, $\pi^- + p \to (N^\ast, \Delta^\ast) \to n + a$, without violating the spontaneously broken chiral symmetry of QCD.

To evaluate the axion emission rate of $\pi^- + p \to n + a$, we need the interactions among the pions, baryons, and axion, especially the axion couplings to nucleons and decuplet baryons.\,\,As mentioned in the previous paragraph, the pion momentum is $\vkpi \simeq m^{}_\pi \ll m_p$ inside SNe.\,\,In other words, the pion momentum is relatively smaller than the proton mass when scattering off the proton.\,\,Such a low-energy pion interacting with a heavy nucleon can be well described by the heavy baryon chiral perturbation theory (HBChPT) proposed in Refs.\,\cite{Jenkins:1990jv,Jen:1991}.\,\,Accordingly, we will adopt the HBChPT to derive the relevant interactions of the process $\pi^- + p \to n + a$ in this paper.\,\,In the HBChPT, the nucleon is almost on shell with a nearly unchanged velocity $v$ when it exchanges some tiny momentum with the pion, and its four-momenta can be divided into $k^{}_N = m^{}_N v + \delta k^{}_\pi$ with $v^2=1$, where $\delta k^{}_\pi$ is a small residual four-momenta coming from the pion.\,\,In this formalism, the power counting expansion of the effective field theory for pions and baryons can be systematic and well behaved.\,\,Also, the effects of higher resonances such as $\Delta(1232)$ decuplet with $I = J = 3/2$ or excited nucleons $N$ with $I = J = 1/2$ can be taken into account in a much better way with systematic power counting rules in the HBChPT, unlike the old-fashioned chiral Lagrangian with baryons.\,\,Further, the advantage of using the HBChPT is that the algebra of the spin operator formalism can be much simpler than that of the gamma matrix formalism when computing the scattering amplitude of the process $\pi^- + p \to n + a$.\,\,We will see this advantage in the later section.

The outline of this paper is as follows.\,\,In the next section, we write down the Lagrangian for the HBChPT and show the interactions of the pions, nucleons, and decuplet baryons.\,\,In Sec.\,\ref{sec:3}, we write down the Lagrangian of the QCD axion and derive the axion interactions to the pions, nucleons, and decuplet baryons.\,\,With the interactions in Sec.\,\ref{sec:2} and Sec.\,\ref{sec:3}, we then compute in Sec.\,\ref{sec:4} the scattering cross section of the process $\pi^- + p \to n + a$ to see the resonance behavior of the cross section due to the $\Delta(1232)$ baryon.\,\,In Sec.\,\ref{sec:5}, we estimate the axion emission rate of the process $\pi^- + p \to n + a$ including the $\Delta(1232)$ resonance contribution in some axion models and discuss its effect on the SN axion emissivity.\,\,We conclude our work in the last section.

\section{Heavy Baryon Chiral Perturbation Theory}\label{sec:2}

In this section, we will write down the chiral Lagrangian density describing the interactions between pions and baryons in the heavy baryon formalism.\,\,In particular, we will show the pion couplings to octet and decuplet baryons and the hadron axial vector currents which are crucial for the $\Delta$ resonance contribution to the axion emission rate of a supernova.\,\,For more detailed discussions of the HBChPT, one can refer to Refs.~\cite{Jenkins:1990jv,Jen:1991,Jenkins:1992pi}.

Firstly, let us write down the lowest order effective chiral Lagrangian containing the
heavy baryon octet $\Bv$ and the meson octet $\bs{\pi}$ as follows~\cite{Jen:1991}\,:
\begin{eqnarray}\label{LBpi}
{\cal L}^{}_{\pi \hs{-0.03cm} B} 
\Eq
\frac{1}{4} f_\pi^2 
\sx{1.1}{\big\langle}
\partial^\mu \bs{\Pi} {}^{} \partial_\mu \bs{\Pi}^\dag 
\sx{1.1}{\big\rangle}
+
b {}^{}{}^{}
\sx{1.1}{\big\langle} {}^{}
\Mq^{} \big({}^{} \bs{\Pi} + \bs{\Pi}^\dag \big)
\sx{1.1}{\big\rangle}
+
i {}^{}{}^{} 
\sx{1.1}{\big\langle} {}^{}{}^{}{}^{}
\OL{\Bv} {}^{} v^\mu {}^{} \Dmu {}^{} \Bv 
\sx{1.1}{\big\rangle}
\nn[0.1cm]
&&
+\,
2D {}^{}{}^{} 
\sx{1.1}{\big\langle} {}^{}{}^{}{}^{}
\OL{\Bv} {}^{} \Sv \big\{ \Amu{}^{}, \Bv \big\} 
\sx{1.1}{\big\rangle}
+
2F{}^{}
\sx{1.1}{\big\langle} {}^{}{}^{}{}^{}
\OL{\Bv} {}^{} \Sv \big[ \Amu{}^{}, \Bv\big] 
\sx{1.1}{\big\rangle}
+
\cdots
~,
\end{eqnarray}
where $\langle\,\cdots\rangle = \tx{tr}(\,\cdots)$ denotes the trace of a matrix,
\begin{eqnarray}
\begin{gathered}
\bs{\Pi}
\,=\,
\xi^2
~,\quad
\xi
\,=\,
\exp \hs{-0.05cm} \bigg( \frac{i\bs{\pi}}{f^{}_\pi} \bigg)
~,\quad
\bs{\pi} 
\,=\,
\frac{1}{\sqrt{2}}
\renewcommand\arraystretch{1.2}
\begin{pmatrix}
\,\frac{1}{\sqrt{2}}{}^{}\pi^0 + \frac{1}{\sqrt{6}}{}^{}\eta & \pi^+ & K^+ \,\\ 
\pi^- & -\frac{1}{\sqrt{2}}{}^{}\pi^0 + \frac{1}{\sqrt{6}}{}^{}\eta & K^0 \,\\ 
K^- & \bar{K}^{}_0 & -\frac{2}{\sqrt{6}}{}^{}\eta  \,\\ 
\end{pmatrix}
~,
\\[0.1cm]
\Bv
\,=\,
\renewcommand\arraystretch{1.2}
\begin{pmatrix}
\,\frac{1}{\sqrt{2}}{}^{}\Sigma^0_v + \frac{1}{\sqrt{6}}{}^{}\Lambda^{}_v & \Sigma^+_v & p^{}_v \,\\ 
\Sigma^-_v & -\frac{1}{\sqrt{2}}{}^{}\Sigma^0_v + \frac{1}{\sqrt{6}}{}^{}\Lambda^{}_v & n^{}_v \,\\ 
\Xi^-_v & \Xi^0_v & -\frac{2}{\sqrt{6}}{}^{}\Lambda^{}_v \,\\ 
\end{pmatrix}
~,\quad
\Dmu {}^{} \Bv
\,=\,
\partial^{}_\mu {}^{} \Bv + \big[{}^{}\Vmu{}^{}, \Bv\big]
~,
\\[0.2cm]
\Vmu
\,=\, 
\frac{1}{2} 
\big({}^{}\xi{}^{} \partial^{}_\mu {}^{} \xi^\dag + \xi^\dag \partial^{}_\mu \xi {}^{}{}^{}\big)
~,\quad
\Amu
\,=\,
\frac{i}{2} 
\big({}^{}\xi{}^{} \partial^{}_\mu {}^{} \xi^\dag - \xi^\dag \partial^{}_\mu \xi {}^{}{}^{}\big)
\end{gathered}
\end{eqnarray}
with $f^{}_\pi \simeq 92.4\,\tx{MeV}$ as the pion decay constant~\cite{Choi:2021ign}, $\Sv = \gamma^5 \big[{}^{}{}^{}\slashed{v}{}^{},\gamma^\mu {}^{} \big]/4$ as the spin operator with $v \cdot S^{}_v = 0$, and $\Mq^{} = \tx{diag}(m^{}_u,m^{}_d,m^{}_s)$ as a diagonal light quark mass matrix which explicitly breaks the global chiral symmetry of the Lagrangian, $\SUL \otimes \SUR$ down to $\tx{SU}(3)_V$.\,\,Under the $\SUL \otimes \SUR$ symmetry, the baryon and meson octets transform as
\begin{eqnarray}
\begin{gathered}
\bs{\Pi} (x) \to\, \UL \bs{\Pi} (x) \,{}^{} \URd
~,\quad 
\xi (x) \to\, \UL {}^{}{}^{} \xi (x) \, \UHd (x) = \UH (x) {}^{}{}^{} \xi (x) \, \URd
~,\quad
\\[0.2cm]
\Bv (x) \to\, \UH (x) {}^{} \Bv (x) {}^{}{}^{}{}^{} {\cal U}_H^\dagger (x)
~,\quad 
\Dmu {}^{} \Bv (x) \to\, \UH (x) \big[ \Dmu {}^{} \Bv (x) \big] {}^{}{}^{}
{\cal U}_H^\dagger (x)
~,
\\[0.2cm]
\Vmu (x) \to\, \UH (x) \Vmu (x) \, \UHd (x) + \UH (x) {}^{} \partial^{}_\mu {}^{} \UHd (x) 
~,\quad
\Amu (x) \to\, \UH (x) {}^{} \Amu (x) \, \UHd (x)
~,\quad
\end{gathered}
\end{eqnarray}
where ${\cal U}^{}_L$ and ${\cal U}^{}_R$ are the group elements of $\SUL$ and $\SUR$, respectively, and ${\cal U}^{}_H(x)  = {\cal U}^{}_H(\xi (x),{\cal U}^{}_L,
{\cal U}^{}_R)$ depending on $x$ via $\xi (x)$ is the group element of hidden local $\tx{SU}(3)^{}_{\hs{-0.03cm}H}$.\,\,Now, to the first order in $\bs{\pi}/f^{}_\pi$, $\xi = \mathbb{I}^{}_{3 \times 3} + i \bs{\pi}/f^{}_\pi{}^{}$,\,\,it follows that $\Vmu = 0$ and $\Amu = \partial^{}_\mu \bs{\pi}/f^{}_\pi$.\,\,Plugging these $\Vmu$ and $\Amu$ into Eq.\,\eqref{LBpi}, we then yield
\begin{eqnarray}\label{LpiB2}
{\cal L}^{}_{\pi \hs{-0.03cm} B} 
\,\supset \,
\frac{2(D+F)}{f^{}_\pi} 
\sx{1.1}{\big\langle} {}^{}{}^{}{}^{} 
\OL{\Bv} {}^{} \Sv \big( \partial^{}_\mu \bs{\pi} \big) \Bv
\sx{1.1}{\big\rangle}
+
\frac{2(D-F)}{f^{}_\pi} 
\sx{1.1}{\big\langle} {}^{}{}^{}{}^{}
\OL{\Bv} {}^{} \Sv \Bv \big( \partial^{}_\mu \bs{\pi} \big) 
\sx{1.1}{\big\rangle}
~,
\end{eqnarray}
from which the interactions of the charged pions and nucleons can be extracted as
\begin{eqnarray}\label{LpiN}
{\cal L}_{{}^{} \pi \hs{-0.02cm} N}
\Eq
\frac{\sqrt{2}{}^{}g^{}_A}{f^{}_\pi}
\sx{1.0}{\big(}\,{}^{}
\OL{p^{}_v} {}^{}{}^{} \Sv {}^{} n^{}_v \partial^{}_\mu \pi^+  
+
\OL{n^{}_v} {}^{}{}^{} \Sv {}^{}{}^{} p^{}_v \partial^{}_\mu \pi^-
\sx{1.0}{\big)}
~,
\end{eqnarray}
where $g^{}_A = D + F \simeq 1.254$~\cite{Vonk:2021sit} is the axial coupling.\,\,Notice that the $D-F$ term in Eq.\,\eqref{LpiB2} does not contribute to the charged pion-nucleon interactions.

Next, we write down the lowest order effective chiral Lagrangian including the interactions between the baryon octet, meson octet, and the spin-3/2 baryon decuplet which is described by a Rarita-Schwinger field $\Tmuu = (\Tmuu)^{}_{ijk}$ with $v \cdot {\cal T}^{}_v = S^{}_v \cdot {\cal T}^{}_v = 0$~\cite{Jen:1991,Jenkins:1992pi,Haidenbauer:2017sws}
\begin{eqnarray}\label{LpiBT}
{\cal L}^{}_{\pi \hs{-0.03cm} B T} 
\Eq
-{}^{}{}^{}i\,
\OL{\displaystyle\big({}^{}\Tmuu{}^{}\big)_{\hs{-0.05cm}ijk}}
{}^{}{}^{} v^\rho {}^{} {\cal D}^{}_{\hs{-0.05cm}\rho}
\big({}^{}\Tmud{}^{}\big)_{\hs{-0.05cm}ijk} 
+
\Delta m^{}_{{}^{}T \hs{-0.03cm} B}
\OL{\displaystyle\big({}^{}\Tmuu{}^{}\big)_{\hs{-0.05cm}ijk}}{}^{}
\big({}^{}\Tmud{}^{}\big)_{\hs{-0.05cm}ijk} 
\nn[0.1cm]
&&+\,
{\cal C} {}^{} \epsilon^{}_{ijk}
\Big[\,
\OL{\displaystyle\big({}^{}\Tmuu{}^{}\big)_{\hs{-0.05cm}i{}^{}\ell m}}
\big(\Amu {}^{}\big)_{\hs{-0.05cm}\ell j}
\big(\Bv\big)_{\hs{-0.05cm}mk}
+
\OL{\big(\Bv\big)_{\hs{-0.05cm}km}}
\big(\Amu {}^{}\big)_{\hs{-0.05cm}j\ell}
\big({}^{}\Tmud{}^{}\big)_{\hs{-0.05cm}i{}^{}\ell m}
\Big]
+
\cdots
~,
\end{eqnarray}
where ${\cal D}_{\hs{-0.03cm}\rho}({}^{}\Tmud{}^{})^{}_{ijk} 
= \partial_{\rho} ({}^{}\Tmud{}^{})^{}_{ijk} + 
({\cal V}_{\hs{-0.03cm}\rho})^\ell_i ({}^{}\Tmud{}^{})^{}_{\ell jk} + 
({\cal V}_{\hs{-0.03cm}\rho})^\ell_j ({}^{}\Tmud{}^{})^{}_{i \ell k} + 
({\cal V}_{\hs{-0.03cm}\rho})^\ell_k ({}^{}\Tmud{}^{})^{}_{ij \ell}{}^{}$, $\Delta m^{}_{{}^{}T \hs{-0.03cm} B} = m^{}_T - m^{}_B$, and ${\cal C} \simeq 3{}^{}g^{}_A/2$~\cite{Haidenbauer:2017sws}.\,\,Under the $\SUL \otimes \SUR$ symmetry, the baryon decuplet transforms as
\begin{eqnarray}
\big({}^{}\Tmuu{}^{}\big)_{\hs{-0.05cm}ijk} 
\to
(\,\UH)^{}_{i{}^{}\ell} \, (\,\UH)^{}_{jm} (\,\UH)^{}_{kn} 
\big({}^{}\Tmuu{}^{}\big)_{\hs{-0.05cm}lmn} 
~,
\end{eqnarray}
with which one can check that Eq.\,\eqref{LpiBT} is invariant under the chiral symmetry.\,\,To explicitly find out the interactions among pions, nucleons, and $\Delta$ baryons, we use the following representation of the $\Delta$ baryons in terms of the above symmetric three-index tensor~\cite{Haidenbauer:2017sws}\,:
\begin{eqnarray}
\big({}^{}\Tmud{}^{}\big)_{\hs{-0.05cm}111} \,=\, \Delta^{++}_{v\mu} ~,\quad
\big({}^{}\Tmud{}^{}\big)_{\hs{-0.05cm}112} \,=\, \frac{1}{\sqrt{3}} {}^{} \Delta^+_{v\mu} ~,\quad
\big({}^{}\Tmud{}^{}\big)_{\hs{-0.05cm}122} \,=\, \frac{1}{\sqrt{3}} {}^{} \Delta^0_{v\mu} ~,\quad
\big({}^{}\Tmud{}^{}\big)_{\hs{-0.05cm}222} \,=\, \Delta^-_{v\mu} ~,\quad
\end{eqnarray}
from which the pion-nucleon-$\Delta$ interactions related to our study are extracted as
\begin{eqnarray}\label{LpiND}
{\cal L}_{{}^{} \pi \hs{-0.02cm} N \hs{-0.03cm} \Delta}
\Eq
\frac{{\cal C}}{\sqrt{6} f^{}_\pi}
\sx{1.0}{\big(}\,
\OL{n^{}_v} {}^{}{}^{} 
\white{\OL{\black{\Delta^+_{v\mu}}}} {}^{} \partial^\mu \pi^- 
+
\OL{\Delta^+_{v\mu}} {}^{}{}^{} n^{}_v {}^{} \partial^\mu \pi^+
-
\OL{p^{}_v} {}^{}{}^{} 
\white{\OL{\black{\Delta^0_{v\mu}}}} {}^{} \partial^\mu \pi^+  
-
\OL{\black{\Delta^0_{v\mu}}} {}^{}{}^{}{}^{} p^{}_v {}^{}{}^{} \partial^\mu \pi^-
\sx{1.0}{\big)}
~.
\end{eqnarray}

Finally, let us write down the hadronic axial vector currents associated with ${\cal L}^{}_{\pi \hs{-0.03cm} B}$ and ${\cal L}^{}_{\pi \hs{-0.03cm} B T} $ invariant under the local $\tx{SU}(3)^{}_{\hs{-0.03cm}H}$ symmetry.\,\,Considering an infinitesimal transformation of the meson field, $\xi \to \UH {}^{}{}^{} \xi \,\URd \to (1+ i {}^{} \epsilon^A t^A) {}^{} \xi$ with $\epsilon^A \to 0{}^{}$, and employing the conserved current in Noether's theorem, one can obtain the corresponding axial vector currents ${\cal J}^{A \mu}$ as~\cite{Jen:1991}
\begin{eqnarray}
{\cal J}^{A \mu}_{\pi \hs{-0.03cm} B}
\Eq
D{}^{}{}^{}
\sx{1.1}{\big\langle}{}^{}{}^{}{}^{}
\OL{\Bv} {}^{} \Sv \big\{ \xi^\dag t^A \xi + \xi {}^{}{}^{} t^A \xi^\dag, \Bv \big\}
\sx{1.1}{\big\rangle}
+
F{}^{}
\sx{1.1}{\big\langle}{}^{}{}^{}{}^{}
\OL{\Bv} {}^{} \Sv \big[{}^{} \xi^\dag t^A \xi + \xi {}^{}{}^{} t^A \xi^\dag, \Bv \big]
\sx{1.1}{\big\rangle}
\nn[0.1cm]
&&
+{}^{}{}^{}
\frac{1}{2} v^\mu
\sx{1.1}{\big\langle}{}^{}{}^{}{}^{}
\OL{\Bv} {}^{} \big[{}^{} \xi^\dag t^A \xi - \xi {}^{}{}^{} t^A \xi^\dag, \Bv\big]
\sx{1.1}{\big\rangle}
~,
\end{eqnarray}
\vs{-0.8cm}
\begin{eqnarray}
\hs{-0.5cm}
{\cal J}^{A \mu}_{\pi \hs{-0.03cm} B T}
\Eq
\frac{{\cal C}}{2}{}^{}{}^{}
\epsilon^{}_{ijk}
\Big[\,
\OL{\displaystyle\big({}^{}\Tmuu{}^{}\big)_{\hs{-0.05cm}i{}^{}\ell m}}
\sx{1.2}{\big(}
\xi^\dag t^A \xi + \xi {}^{}{}^{} t^A \xi^\dag 
\sx{1.2}{\big)}_{\hs{-0.08cm}\ell j}
\big(\Bv\big)_{\hs{-0.05cm}mk}
+
\OL{\big(\Bv\big)_{\hs{-0.05cm}km}}
\sx{1.2}{\big(}
\xi^\dag t^A \xi + \xi {}^{}{}^{} t^A \xi^\dag 
\sx{1.2}{\big)}_{\hs{-0.08cm}\ell j}
\big({}^{}\Tmud{}^{}\big)_{\hs{-0.05cm}i{}^{}\ell m}
\Big]
~,
\end{eqnarray}
where $t^A\,(A=1,2,\cdots,8)$ are the Gell-Mann matrices with the normalization $\big\langle t^A t^B \big\rangle = \delta^{AB}/2$. We will utilize these hadron axial vector currents to derive the interactions among the axion, nucleons and decuplet baryons in the next section.

\section{Axion couplings to baryons and mesons}\label{sec:3}

In this section, we will show the derivation of the interactions between the QCD axion and baryons and mesons, particularly the axion coupling to decuplet baryons, in the HBChPT.\,\,We first write down the effective Lagrangian of the QCD axion in two representative axion models, the Kim-Shifman-Vainshtein-Zakharov (KSVZ) model \cite{Kim:1979if,Shifman:1979if} and the Dine-Fischler-Srednicki-Zhitnitsky (DFSZ) model \cite{Zhitnitsky:1980tq,Dine:1981rt}, and perform a chiral transformation on the light quark fields to eliminate the axion-gluon interaction as usual.\,\,In this quark field basis, we can then match the couplings of the axion to quarks and gluons above the QCD confinement scale onto that of the axion to baryons and mesons below the QCD confinement scale.\footnote{A more detailed discussion of this procedure can be found in Ref.\,\cite{Georgi:1986df}.}

The most general effective Lagrangian of the QCD axion, $a(x)$, with the light quark fields, $q = (u,d,s){}^\tf{T}$, below the PQ and EW breaking scales and above the scale of QCD confinement can be expressed at leading order in $a/f^{}_a$ (here we omit the axion interaction with photons as it is irreverent to our study) as 
\begin{eqnarray}\label{Laqg}
{\cal L}^{}_{aqg}
=\,
\frac{1}{2}{}^{} \partial_\mu a {}^{}{}^{} \partial^\mu a 
+ 
\frac{g_s^2}{32\pi^2} \frac{a}{f^{}_a} {}^{}{}^{} G^{c}_{\mu\nu} \widetilde{G}^{c\mu\nu} 
+
\OL{q} {}^{}{}^{}i{}^{} \gamma^\mu \partial^{}_\mu q \,
-
\big(\,
\OL{\qL} {}^{}{}^{} \Mq^{} {}^{}{}^{} \qR + \tx{h.c.}
\big)
+
\frac{\partial_\mu a}{2 f^{}_a} {}^{}{}^{} \OL{q} {}^{}{}^{} \gamma^\mu \gamma^5
\Xq {}^{} q 
~,
\end{eqnarray}
where $f^{}_a$ is the axion decay constant, $g^{}_s$ is the gauge coupling of the strong interaction, $G^c_{\mu\nu}$ with $c$ being the color index is the gluon field strength tensor and $\widetilde{G}^{c\mu\nu} = \epsilon^{\mu\nu\rho\sigma} G^c_{\rho\sigma}/2$ with $\epsilon^{0123} = +1$ is its dual tensor, $q^{}_{L,R} = P^{}_{L,R} {}^{}{}^{} q$ with $P^{}_{L,R} = (1 \mp \gamma^5)/2$, and $\Mq^{}$ is the quark mass matrix defined in the previous section, and the last term in Eq.\,\eqref{Laqg} denotes the axion derivative interactions with the quark axial vector currents with $\Xq = \tx{diag}(X_u, X_d, X_s)$ being a coupling matrix depending on a UV model above the PQ symmetry breaking scale.\,\,Typically, one introduces an SM-singlet complex scalar field $\Phi \sim (\bs{1},\bs{1})_0$ with a PQ charge in these UV models.\,\,After the PQ symmetry breaking, the phase of $\Phi$ is then identified as the axion which couples to the SM gluons due to the QCD anomaly.\,\,In the KSVZ model, the QCD anomaly is realized by introducing a heavy vector-like fermion ${\cal Q} = \cQL + \cQR \sim (\bs{3},\bs{1})_0$ which couples to the PQ scalar $\Phi$ via the Yukawa interaction, $y^{}_Q \Phi {}^{}{}^{} \OL{\cQL} \white{\OL{\black{\cQR}}} + \tx{H.c.}$, where $\Phi \to e^{i \qPQ} \Phi, \cQL \to e^{i \qPQ/2} \cQL$, and $\cQR \to e^{-i \qPQ/2} \cQR$ under the PQ symmetry.\,\,Since only $\Phi$ and ${\cal Q}$ have the PQ charges, implying that the axion interacts with the SM quark fields radiatively~\cite{Choi:2021kuy}, $\Xq = 0$ at tree level in the KSVZ model.\,\,In the DFSZ model, the QCD anomaly is induced by assuming two Higgs doublets $\Hu$ and $\Hd$ which couple to the SM quarks, $\QL, \UR{}^{}, \tx{and}\,\DR$ via the Yukawa interactions, $\OL{\QL}\big( {\cal Y}^{}_u \widetilde{H}^{}_{\hs{-0.03cm}u} {}^{} \UR + {\cal Y}^{}_d {}^{} \Hd {}^{} \DR {}^{}\big) + \tx{H.c.}$, and the PQ scalar $\Phi$ couples to these two Higgs doublets via the terms in the scalar potential, e.g.,\,$H_{\hs{-0.03cm}u}^\dagger \Hd (\Phi^\ast)^2$, where $\Phi \to e^{i \qPQ} \Phi, \Hu \to e^{-i \qPQ} \Hu{}^{}, \Hd \to e^{i \qPQ} \Hd {}^{}, \QL \to \QL{}^{}, \UR \to e^{-i \qPQ} \UR{}^{}$, and $\DR \to e^{-i \qPQ} \DR$ under the PQ symmetry.\footnote{The DFSZ model can further classify into the DFSZ-I and DFSZ-II models, in which the leptophilic Yukawa interactions are $\OL{\LL} {\cal Y}^{}_e {}^{} \Hd \ER$ and $\OL{\LL} {\cal Y}^{}_e {}^{} \Hu \ER$, respectively, with $\LL$ and $\ER$ being the SM lepton fields.\,\,However, since the Higgs doublet couplings to the SM quarks are the same in these two models and the supernova axion emission are hadronic processes, we do not distinguish these two models in our calculations.}\,\,After the PQ and the EW symmetry breaking, the axion field which is one of the linear superpositions of the $C\hs{-0.03cm}P$-odd scalars in $\Hu, \Hd$ and $\Phi$ can couple to the SM quarks at tree level.\footnote{A detailed calculation of the DFSZ axion couplings to the SM fermions can be found in a recent paper~\cite{Sun:2020iim}.}\,\,Here we summarize the axion couplings to the light quarks at tree level in the KSVZ and DFSZ models below~\cite{DiLuzio:2020wdo,Ferreira:2020bpb}\,:
\begin{eqnarray}\label{KD}
\tx{KSVZ model\,\,:\,\,} X_u = X_d {}^{} = X_s = 0 
~~;~~
\tx{DFSZ model\,\,:\,\,} 
X_u = \frac{\cos^2 \hs{-0.07cm} \beta}{N_g}
 ~,~~
X_d {}^{} = X_s = \frac{\sin^2 \hs{-0.07cm} \beta}{N_g}
~,
\end{eqnarray}
where $N_g = 3$ is the number of the SM fermion generations, and $\tan \beta = \upsilon^{}_u/\upsilon^{}_d$ with $\upsilon^{}_u$ and $\upsilon^{}_d$ being the vacuum expectation values of $\Hu$ and $\Hd$, respectively.

To compute the axion couplings to baryons and mesons below the scale of QCD confinement, we can first remove the axion-gluon interaction explicitly by the following chiral transformation on the light quark fields as~\cite{Georgi:1986df}
\begin{eqnarray}\label{qtrans}
q
\,\to\,
{\cal R}^{}_a {}^{} q
\,=\,
\exp \hs{-0.05cm} 
\bigg(\hs{-0.05cm}{-}{}^{}{}^{} i \gamma^5 \frac{a}{2f^{}_a} {}^{} \Qa\bigg) q
~,\quad
\langle \Qa {}^{} \rangle \,=\, 1 
~,
\end{eqnarray}
where $\Qa$ is a real 3 by 3 matrix acting on the quark flavor space.\footnote{With the convention of $\epsilon^{0123} = +1$, the functional measure in the quark field functional integration gives~\cite{Peskin:1995ev}
\begin{eqnarray}\label{Dqtrans}
\int \hs{-0.08cm} {\cal D}q {\cal D}\bar{q} 
\to 
\int \hs{-0.08cm} {\cal D}q {\cal D}\bar{q} {}^{} 
\exp \hs{-0.05cm} 
\bigg[ i \int \hs{-0.08cm} \dd^4 x 
\bigg(\hs{-0.05cm}
{-}{}^{}{}^{}\frac{g_s^2}{32\pi^2} \frac{a}{f^{}_a} {}^{}{}^{} 
G^c_{\mu\nu} \widetilde{G}^{c\mu\nu} \langle \Qa {}^{} \rangle
\hs{-0.05cm}\bigg)
\bigg]
\end{eqnarray}
under the chiral transformation in \eqref{qtrans}, where we take $\langle \Qa {}^{} \rangle = 1$ to cancel the axion-gluon interaction in \eqref{Laqg}.}\,\,To avoid the axion-$\pi^0$ mass mixing, the convenient choice of $\Qa$ is given by\footnote{Even with this customary choice of $\Qa$, there is still an axion-$\pi^0$ kinetic mixing in the Lagrangian.\,\,However, since the strength of this kinetic mixing $\epsilon_{a\pi} \sim {\cal O}(m^{}_a/m^{}_\pi) \ll 1$, it is usually ignored in the literature~\cite{DiLuzio:2020wdo}.}~\cite{Georgi:1986df}
\begin{eqnarray}
\Qa 
\,=\, 
\frac{{\cal M}^{-1}_q}{\mathrm{tr}{}^{}\big({\cal M}^{-1}_q\big)}
\,=\,
\frac{m^{}_u m^{}_d {}^{} m^{}_s}{m^{}_u m^{}_d  + m^{}_u  m^{}_s +m^{}_d {}^{} m^{}_s}
{}^{}{}^{}\mathrm{diag}\bigg(\frac{1}{m^{}_u},\frac{1}{m^{}_d},\frac{1}{m^{}_s}\bigg)
~.
\end{eqnarray}
On the other hand, under this chiral transformation, the quark kinetic term in \eqref{Laqg} is shifted as
\begin{eqnarray}\label{qkin}
\OL{q} {}^{}{}^{}i{}^{} \gamma^\mu \partial^{}_\mu q
\,\to\,
\overline{q} {}^{}{}^{}i{}^{} \gamma^\mu \partial^{}_\mu q 
+
\frac{\partial^{}_\mu a}{2f^{}_a} {}^{}{}^{}
\OL{q} {}^{}{}^{} \gamma^\mu \gamma^5 \Qa {}^{} q
+ 
{\cal O}\bigg(\frac{a^2}{f_a^2}\bigg)
~,
\end{eqnarray}
while the light quark mass term becomes
\begin{eqnarray}\label{qmass}
\OL{\qL} {}^{}{}^{} \Mq^{} {}^{}{}^{} \qR
\,\to\,
\OL{\qL} {}^{}{}^{} \Ma^{} {}^{}{}^{} \qR
~,\quad
\OL{\qR} {}^{}{}^{} \Mq^{} {}^{}{}^{} \qL
\,\to\,
\OL{\qR} {}^{}{}^{} {\cal M}_a^\dagger {}^{}{}^{} \qL
~,
\end{eqnarray}
where $\Ma^{} \,\equiv\, {\cal R}^{}_a {}^{} \Mq^{} {\cal R}^{}_a$, and up to the second order in $a/f^{}_a$ we have 
\begin{eqnarray}\label{Ma}
\Ma^{}
\,=\,
\Mq^{}
- 
i{}^{}\frac{a}{2f^{}_a} \big\{\Mq^{},\Qa\big\} 
- 
\frac{a^2}{8f_a^2} 
\sx{1.2}{\big\{} \hs{-0.08cm} 
\big\{\Mq^{}, \Qa\big\}, \Qa 
\hs{-0.05cm} \sx{1.2}{\big\}}
+
{\cal O}\bigg(\frac{a^3}{f_a^3}\bigg)
~.
\end{eqnarray}
With Eqs.\,\eqref{qtrans}, \eqref{qkin}, and \eqref{qmass}, the resulting Lagrangian with only the axion and quark fields is 
\begin{eqnarray}\label{Laq}
{\cal L}^{}_{aq} 
\,=\,
\frac{1}{2} \partial_\mu a {}^{}{}^{} \partial^\mu a 
+ 
\OL{q} {}^{}{}^{}i{}^{} \gamma^\mu \partial^{}_\mu q 
+
\sx{1.1}{\big\langle} {}^{} 
\Ma^{} {}^{} \qR {}^{}{}^{} \OL{\qL} 
+ 
{\cal M}^\dag_a {}^{} \qL {}^{} \OL{\qR} \, 
\sx{1.1}{\big\rangle}
+
\frac{\partial_\mu a}{f^{}_a} {}^{} 
\sx{1.1}{\big\langle} 
\big( \Xq + \Qa \big) {}^{} \hat{t}^A 
\sx{1.1}{\big\rangle} {}^{} 
{\cal J}^{A \mu}_q
~,
\end{eqnarray}
where $\displaystyle \big\{ \hat{t}^A \big\} = \big\{ t^A \big\} \cup \big\{ t^0 \big\}$ with $t^0 = \mathbb{I}^{}_ {3\times 3}/\sqrt{6}$ and $\big\langle \hat{t}^A\hat{t}^B \big\rangle = \delta^{AB}/2$ and ${\cal J}^{A \mu}_q = \OL{q} {}^{}{}^{} \gamma^\mu \gamma^5 \hat{t}^A q$ are the quark axial vector currents.\,\,For the last term in the above expression, we have applied the relation $M_{3\times3} = 2\langle M_{3\times3}\,\hat{t}^A \rangle {}^{} \hat{t}^A$ for any 3 by 3 Hermitian matrix $M_{3\times3}$.\,\,Our next step is to replace the light quark fields in Eq.\,\eqref{Laq} with the corresponding hadron fields in the HBChPT.

First, we can replace the $\qL {}^{} \OL{\qR}$ in the third term of Eq.\,\eqref{Laq} with the $\bs{\Pi}$ in Eq.\,\eqref{LBpi} since both have the same transformation properties, $\UL (\qL {}^{} \OL{\qR} {}^{}) \, \URd \sim \UL \bs{\Pi} \,{}^{} \URd$.\,\,With the correct mass dimension, we can write down
\begin{eqnarray}\label{Lapi}
{\cal L}_{a\pi}
\,=\,
\frac{1}{2} f_\pi^2 B^{}_0
\sx{1.1}{\big\langle} {}^{} 
\Ma^{} {}^{} \bs{\Pi}^\dagger
+ 
{\cal M}^\dag_a {}^{} \bs{\Pi} {}^{}{}^{} 
\sx{1.1}{\big\rangle}
~,
\end{eqnarray}
where $B^{}_0$ is determined by the pion mass.\,\,Plugging Eq.\,\eqref{Ma} into Eq.\,\eqref{Lapi}, to the first order in $\bs{\pi}/f^{}_\pi$, one can show that the mass mixing of the axion and $\pi^0$ is automatically eliminated with the choice of $\Qa$ given in Eq.\,\eqref{qtrans}.\,\,On the other hand, the mass of the axion can be expressed in terms of the light quark masses and the pion mass $m^{}_\pi$ as
\begin{eqnarray}
m^{}_a 
\,=\, \sqrt{\frac{z}{(1+z)(1+z+w)}}{}^{}{}^{}\frac{f^{}_\pi m^{}_\pi}{f^{}_a}
\,\simeq\,
6 \,\tx{meV} \bigg(\frac{10^9\,\tx{GeV}}{f^{}_a}\bigg)
~,
\end{eqnarray}
where $m^{}_\pi = \sqrt{B^{}_0(m^{}_u+m^{}_d)} \simeq 139.57\,\tx{MeV}$~\cite{ParticleDataGroup:2022pth}, and $z \equiv m^{}_u/m^{}_d \simeq 0.485$, and $w \equiv m^{}_u/m^{}_s \simeq 0.025$~\cite{Vonk:2021sit}.\,\,In the following sections, we will assume that the axion is massless in our calculations since $m^{}_a \ll m^{}_\pi$ with the typical values of $f^{}_a$ (we will take $f^{}_a = 10^9\,\tx{GeV}$ throughout this paper for our numerical calculations).

Similarly, we can replace the axial vector currents of the light quark fields in Eq.\,\eqref{Laq} with those of the hadron fields in Eq.\,\eqref{LBpi} as follows~\cite{Georgi:1986df}\,:
\begin{eqnarray}\label{LapiB}
{\cal L}^{}_{a \pi \hs{-0.03cm} B}
\Eq
\frac{\partial_\mu a}{f^{}_a} 
\bigg[
\sx{1.1}{\big\langle} 
\big( \Xq + \Qa \big) {}^{} t^A 
\sx{1.1}{\big\rangle} {}^{} 
{\cal J}^{A \mu}_{\pi \hs{-0.03cm} B}
+
\frac{1}{3} S{}^{}{}^{}
\sx{1.1}{\big\langle} 
\Xq + \Qa
\sx{1.1}{\big\rangle} {}^{} 
{\cal J}^{0 \mu}_{\pi \hs{-0.03cm} B}
\bigg]
~,
\end{eqnarray}
where ${\cal J}^{0 \mu}_{\pi \hs{-0.03cm} B} = \sx{1.1}{\big\langle}{}^{}{}^{}{}^{}
\OL{\Bv} {}^{} \Sv \Bv \sx{1.1}{\big\rangle}$ is an isosinglet axial vector current, and
\begin{eqnarray}\label{LaBT}
{\cal L}^{}_{a\pi \hs{-0.03cm} B T}
\Eq
\frac{\partial_\mu a}{f^{}_a} 
\sx{1.1}{\big\langle} 
\big( \Xq + \Qa \big) {}^{} t^A 
\sx{1.1}{\big\rangle} {}^{} 
{\cal J}^{A \mu}_{\pi \hs{-0.03cm} B T}
~,
\end{eqnarray}
which is written down for the first time in this study.\,\,Notice that there is no isosinglet axial vector current including the decuplet baryons since $\epsilon^{}_{ijk} \OL{\displaystyle\big({}^{}\Tmuu{}^{}\big)_{\hs{-0.05cm}ij m}}
\big(\Bv\big)_{\hs{-0.05cm}mk} = 0{}^{}$.\,\,From Eq.\,\eqref{LapiB}, we can obtain the interactions between the axion, pions, and nucleons.\,\,However, they have been derived a number of times in the literature~\cite{Vonk:2021sit,Choi:2021ign,GrillidiCortona:2015jxo}; thus we do not go into the detail of their derivations in this paper.\,\,Here we simply write down these interactions in the HBChPT as\footnote{This can be done by using the identities $\OL{\Bv} {}^{} \gamma^\mu \Bv = v^\mu {}^{} \OL{\Bv} {}^{} \Bv$ and $\OL{\Bv} {}^{} \gamma^\mu \gamma^5 \Bv = 2 {}^{}{}^{} \OL{\Bv} {}^{} S_v^\mu \Bv$~\cite{Jenkins:1990jv}.}
\begin{eqnarray}\label{LapiN}
{\cal L}_{{}^{} a \pi \hs{-0.02cm} N}
\Eq
\frac{\partial_\mu a}{f^{}_a}
\bigg[{}^{}
C^{}_{ap} \,\OL{p^{}_v} {}^{}{}^{} \Sv {}^{} p^{}_v 
+ 
C^{}_{an} {}^{}{}^{} \OL{n^{}_v} {}^{}{}^{} \Sv {}^{} n^{}_v 
+
\frac{i}{2f^{}_\pi}{}^{}\CapiN^{}
\big({}^{}
\pi^+ {}^{} \OL{p^{}_v} {}^{}{}^{} v^\mu {}^{} n^{}_v -
\pi^- {}^{} \OL{n^{}_v} {}^{}{}^{} v^\mu {}^{} p^{}_v {}^{}
\big)
\bigg]
~,
\end{eqnarray}
where the axion couplings to the charged pions and nucleons are given by
\begin{eqnarray}
C^{}_{ap} 
\Eq 
X^{}_u  {}^{} \Delta u + X^{}_d {}^{} \Delta d + X^{}_s {}^{} \Delta s
+
\frac{\Delta u + z \Delta d + w \Delta s}{1+z+w} 
~,\quad
\\[0.1cm]
C^{}_{an} 
\Eq
X^{}_d {}^{} \Delta u + X^{}_u {}^{} \Delta d + X^{}_s {}^{} \Delta s 
+
\frac{z \Delta u + \Delta d + w \Delta s}{1+z+w} 
~,\quad
\\[0.1cm]
\CapiN^{}
\Eq
\frac{1}{\sqrt{2}}
\bigg(X^{}_u - X^{}_d+\frac{1-z}{1+z+w}{}^{}{}^{}\bigg)
\,=\,
\frac{C^{}_{ap} - C^{}_{an}}{\sqrt{2} {}^{}{}^{}g^{}_A}
\label{CapiN}
~.
\end{eqnarray}
In these axion couplings, $\Delta u = 0.847, \Delta d = -{}^{}{}^{}0.407$, and $\Delta s = -{}^{}{}^{}0.035$ are the nucleon matrix elements defined by $\langle {}^{}{}^{}p| {}^{}{}^{} \OL{q} {}^{}{}^{} \Sv q \big|p \rangle = s^\mu \Delta q/2$ with $s^\mu$ being the proton spin~\cite{Vonk:2021sit}.\,\,Notice that there is a contact interaction for $a$-$\pi$-$N$, the $\CapiN^{}$ term in Eq.\,\eqref{LapiN}, which was largely ignored in the literature and should be present in order to respect the spontaneous chiral symmetry breaking of QCD.\,\,Also, its relevance in the axion emission from the SNe was noted in Ref.\,\cite{Choi:2021ign}.

On the other hand, one can extract the interactions of the axion, nucleons, and $\Delta$ decuplet baryons from Eq.\,\eqref{LaBT} as
\begin{eqnarray}\label{LaND}
{\cal L}_{{}^{} a N \hs{-0.03cm} \Delta}
\Eq
\frac{\partial_\mu a}{2f^{}_a}
\Big[{}^{}{}^{}
C^{}_{a p \Delta} 
\sx{1.1}{\big(}{}^{}{}^{}{}^{}
\OL{p^{}_v} {}^{}{}^{} \white{\OL{\black{\Delta^+_\mu}}} 
+ 
\OL{\Delta^+_\mu} {}^{} p^{}_v 
\sx{1.1}{\big)}
+
C^{}_{a n \Delta} 
\sx{1.1}{\big(}{}^{}{}^{}
\OL{n^{}_v} {}^{}{}^{}  \white{\OL{\black{\Delta^0_\mu}}} 
+ 
\OL{\Delta^0_\mu} {}^{} n^{}_v
\sx{1.1}{\big)}
\Big]
~,
\end{eqnarray}
where the axion couplings to the nucleons and $\Delta$ baryons are given by
\begin{eqnarray}
C^{}_{a p \Delta} \Eq C^{}_{a n \Delta}
\,\equiv\, 
\CaND
\,=\,
-\frac{{\cal C}}{\sqrt{3}} {}^{}
\bigg(X^{}_u-X^{}_d+\frac{1-z}{1+z+w}{}^{}{}^{}\bigg)
\,=\,
-{}^{}\frac{\sqrt{3}}{2}\big(C^{}_{ap} - C^{}_{an}{}^{}\big)
\label{CaND}
~.
\end{eqnarray}
Note that this interaction Lagrangian describing $\Delta(1232) \to n + a$ is derived for the first time in the HBChPT.\,\,We shall utilize Eqs.\,\eqref{LpiN} and \eqref{LaND} and the corresponding couplings in order to calculate the SN axion emission rate from the underlying process $\pi^- + p \to \Delta(1232) \to n+a$.

Notice that in our calculation of the axion to hadron couplings, the relative sign between the SM and new physics contributions is opposite to most of the literature~\cite{Vonk:2021sit,Choi:2021ign,GrillidiCortona:2015jxo}.\footnote{Our relative sign of the SM and new physics contributions in the axion-hadron couplings agree with Ref.\,\cite{Georgi:1986df}.}\,\,This relative sign is corresponding to the one between $\Xq$ and $\Qa$ in Eq.\,\eqref{Laq} which originates from the sign in the exponent of the chiral transformation in Eq.\,\eqref{qtrans} and is associated with the convention of $\epsilon^{0123} = +1$.\,\,If one adopts $\epsilon^{0123} = -1$, the prefactor signs of the $a{}^{}G\widetilde{G}$ terms in Eqs.\,\eqref{Laqg} and \eqref{Dqtrans} are both flipped, which keeps the elimination of the $a{}^{}G\widetilde{G}$ term in Eq.\,\eqref{Laqg}, while the sign in the exponent of the chiral transformation in Eq.\,\eqref{qtrans} remains unchanged.\,\,That is to say, the convention of the Levi-Civita tensor has nothing to do with this relative sign.

Finally, one can also note that the $\CapiN$ and $\CaND$ are not independent parameters as they can be expressed in terms of $C^{}_{ap} - C^{}_{an}$ as shown in Eqs.\,\eqref{CapiN} and \eqref{CaND}, respectively.\,\,The values of these axion-hadron couplings are fixed in the KSVZ model and only vary with $\beta$ in the DFSZ model.\,\,With Eq.\,\eqref{KD} and the above numerical inputs, we obtain\footnote{Here we have ignored the heavy quark $(c,b,t)$ contributions to the axion-hadron couplings.}
\begin{eqnarray}
C^{}_{ap} 
\Eq
\begin{cases}
+\,0.430                                                      & \tx{KSVZ model} \\[0.1cm]
+\,0.712 - 0.430 \sin^2 \hs{-0.05cm} \beta &\tx{DFSZ model} \\    
\end{cases}
~,
\\[0.15cm]
C^{}_{an} 
\Eq
\begin{cases}
+\,0.002                                                      & \tx{KSVZ model} \\[0.1cm]
-\,0.134 + 0.406 \sin^2 \hs{-0.05cm} \beta &\tx{DFSZ model} \\    
\end{cases}
~,
\\[0.15cm]
\CapiN
\Eq
\begin{cases}
+\,0.241                                                      & \tx{KSVZ model} \\[0.1cm]
+\,0.477 - 0.471 \sin^2 \hs{-0.05cm} \beta &\tx{DFSZ model} \\    
\end{cases}
\label{CapiNn}
~,
\\[0.15cm]
\CaND
\Eq
\begin{cases}   
-\,0.370                                                       & \tx{KSVZ model} \\[0.1cm]
-\,0.732 + 0.724 \sin^2 \hs{-0.05cm} \beta &\tx{DFSZ model} \\ 
\end{cases}
\label{CaNDn}
~.
\end{eqnarray}
In the later section, we will use these couplings of the axion and hadrons, especially the axion-nucleon-$\Delta$ couplings, to evaluate the supernova energy loss rate induced by the axion emission process $\pi^- + p \to n + a$.\,\,On top of that, we will discuss the effect of the $\Delta$ resonance on the supernova axion emission rate compared with the case without the $\Delta$ resonance.

\section{Scattering cross section of $\normalsize{\bs{\pi^- + p \to n + a}}$}\label{sec:4}

\begin{figure}[t!]
\centering
\hs{0.0cm}
\includegraphics[width=0.31\textwidth]{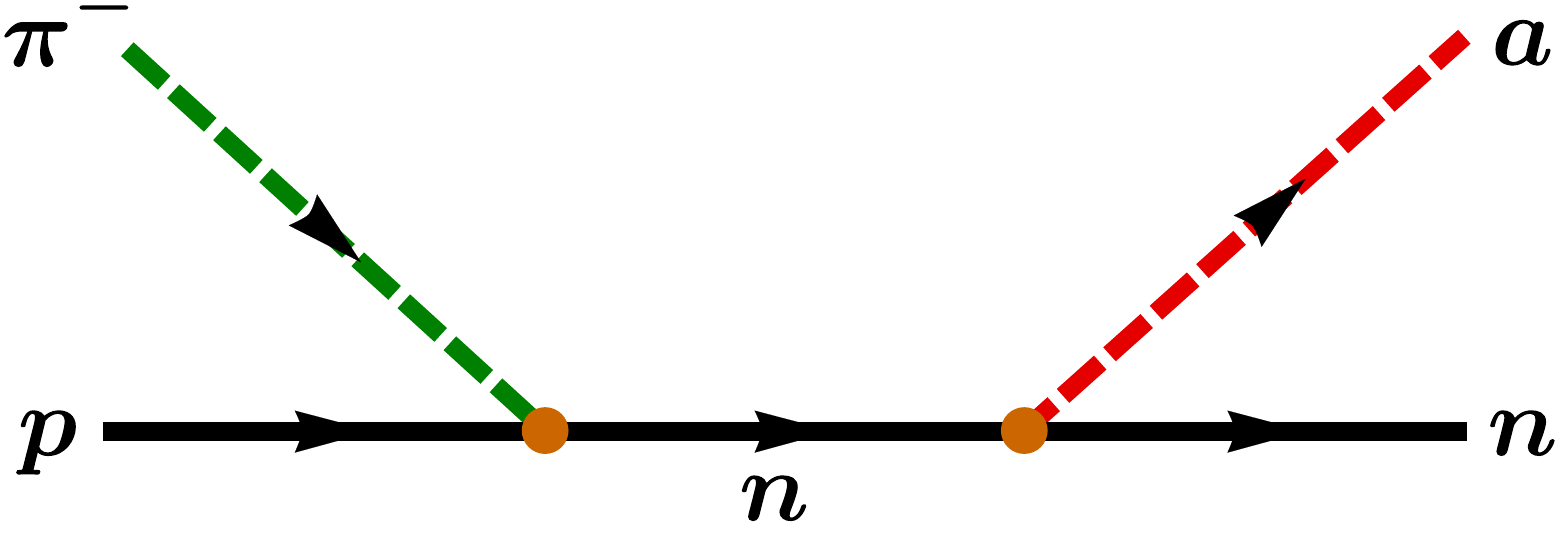}
\hs{0.2cm}
\includegraphics[width=0.31\textwidth]{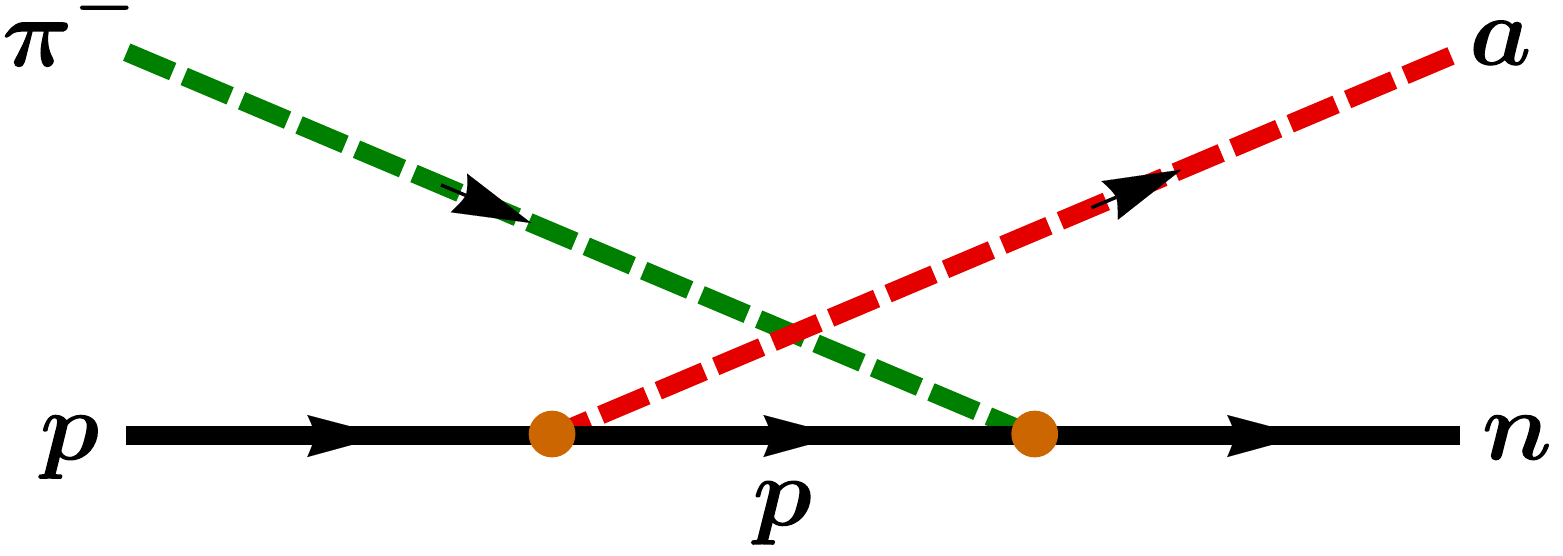}
\hs{0.2cm}
\includegraphics[width=0.31\textwidth]{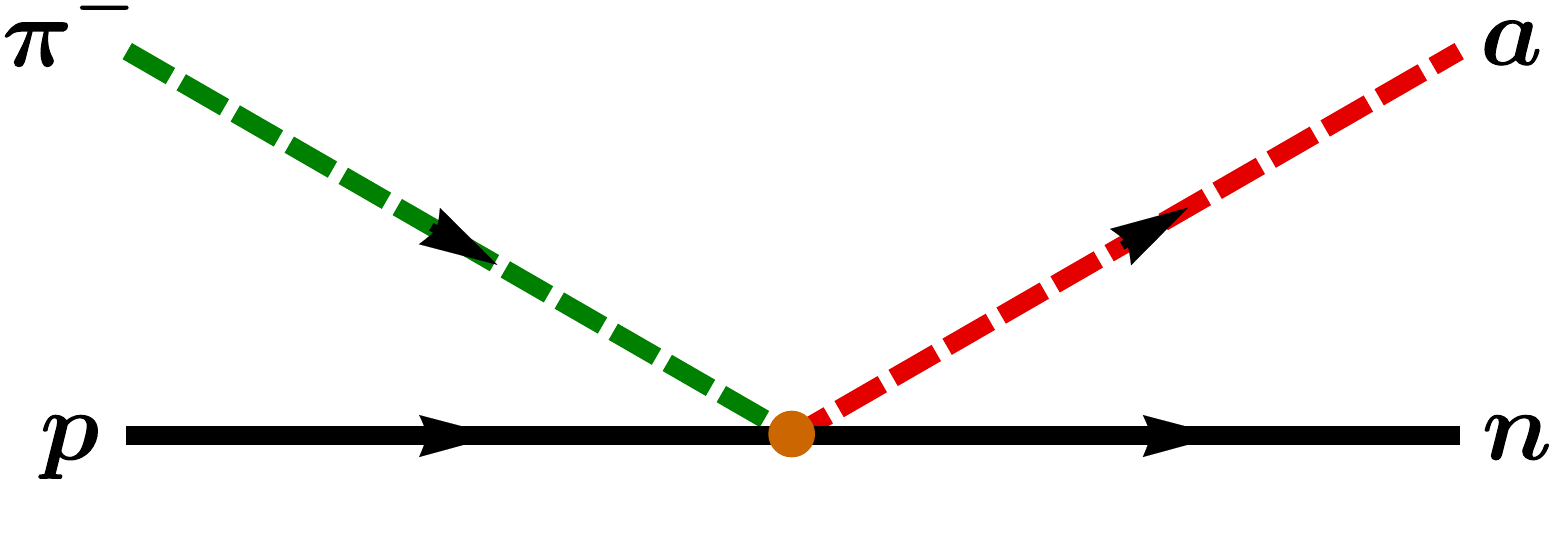}
\\[0.5cm]
\includegraphics[width=0.31\textwidth]{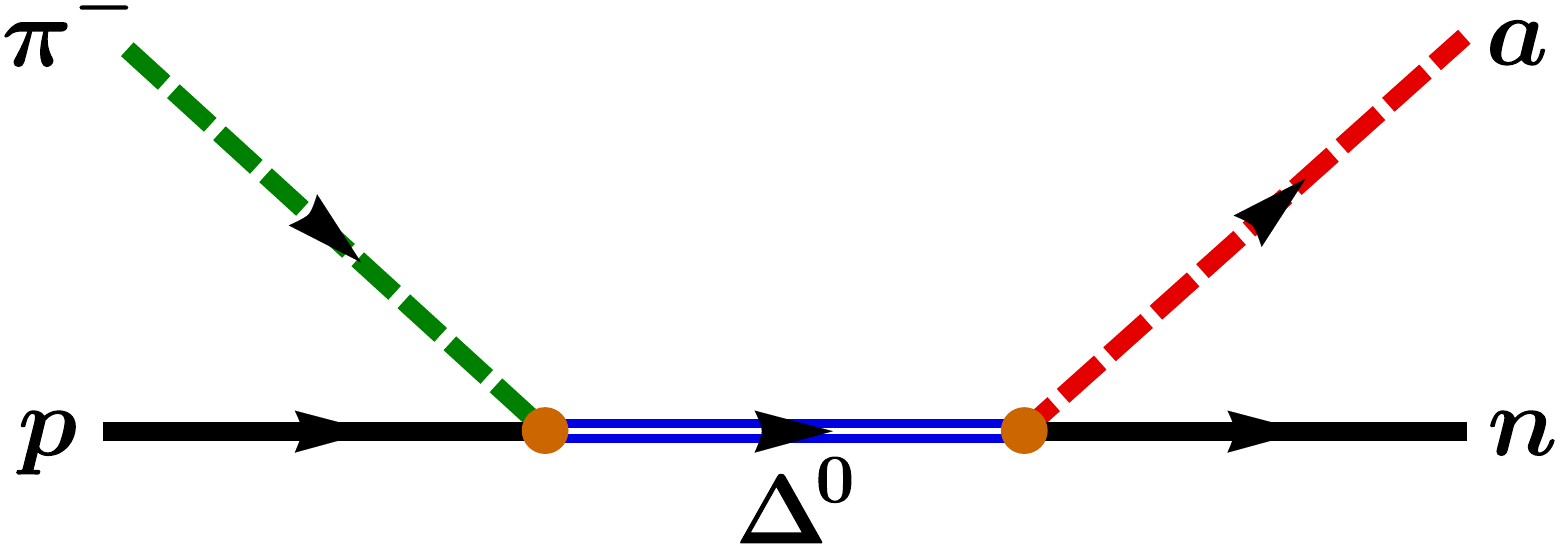}
\hs{0.2cm}
\includegraphics[width=0.31\textwidth]{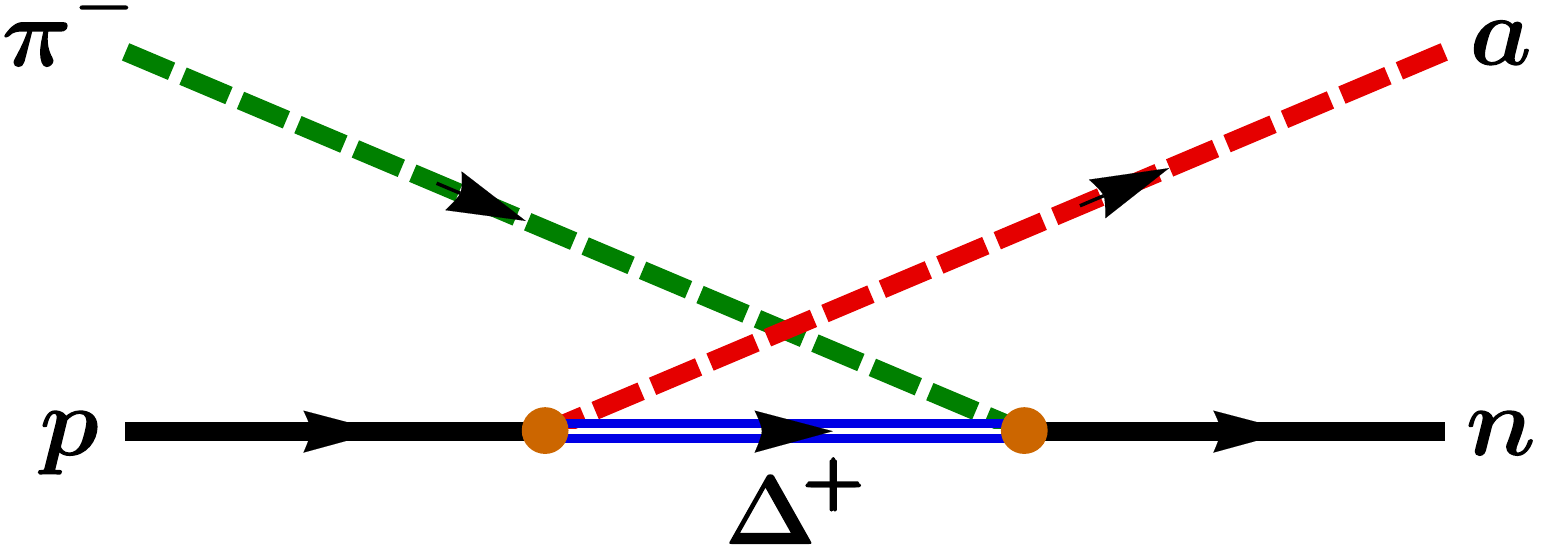}
\vs{-0.0cm}
\caption{Feynman diagrams for $\pi^- + p \to n + a$ with the $\Delta$ baryon contributions.}
\label{fig:pipan}
\vs{0.0cm}
\end{figure}

Before evaluating the supernova axion emission rate, let us first see the resonance behavior in the cross section of the scattering process $\pi^- + p \to n + a$ due to the $\Delta(1232)$ baryon.\,\,With the interactions in Eqs.\,\eqref{LpiN}, \eqref{LpiND}, \eqref{LapiN}, and \eqref{LaND}, the Feynman diagrams of the scattering process $\pi^- + p \to n + a$ are depicted in Fig.\,\ref{fig:pipan}, and the corresponding squared matrix element averaged over the initial spin of the proton is given by\footnote{Here we have normalized the matrix element in the nonrelativistic limit to the one in the relativistic limit by ${\cal M}_{\pi^- p \to n a} = 2{}^{}m^{}_N ({\cal M}_{\pi^- p \to n a})^{}_\tf{NR}$~\cite{Fan:2010gt}.}
\begin{eqnarray}
\OL{\big|{\cal M}_{\pi^- p \to n a}\big|^2}
\,=\,
\frac{2{}^{}m_N^2}{f_\pi^2 f_a^2}
\big\langle P^{}_+ \Omega^\dagger P^{}_+ \Omega {}^{}\big\rangle
~,
\end{eqnarray}
where $m^{}_N = (m^{}_n+m^{}_p)/2 \simeq 938.9\,\text{MeV}$ is the averaged nucleon mass, $P^{}_+ = \tx{diag}(1,1,0,0)$, and
\begin{eqnarray}
\Omega
\Eq
\frac{\sqrt{2}{}^{}g^{}_A \vkpi \vka}{4\Epi}
\sx{1.1}{\big(}
C^{}_{ap} {}^{} \Theta
-
C^{}_{an}         \Theta^\dagger {}^{}
\sx{1.1}{\big)}
+
\frac{\CapiN \vka }{2} {}^{}
{}^{}{}^{} \mathbb{I}^{}_{4\times 4}
\nn[0.1cm]
&&
+{}^{}{}^{}
\frac{{\cal C} {}^{} \vkpi \vka}{6\sqrt{6}}
\Bigg[
\frac{C^{}_{a n \Delta}
\sx{1.1}{\big(}{}^{}
3{}^{}{}^{}\tx{cos}{}^{}\theta{}^{}{}^{}{}^{}
\mathbb{I}^{}_{4\times 4} - \Theta^\dagger 
\sx{1.1}{\big)}}
{E^{}_\pi - \Delta m + i {}^{} \Gamma_\Delta/2}
+
\frac{C^{}_{a p \Delta}
\sx{1.1}{\big(}{}^{}
3{}^{}{}^{}\tx{cos}{}^{}\theta{}^{}{}^{}{}^{}
\mathbb{I}^{}_{4\times 4} - \Theta
\sx{1.1}{\big)}}
{E^{}_\pi + \Delta m - i {}^{} \Gamma_\Delta/2}
\Bigg]
~,
\end{eqnarray}
\newline
where $\Theta = \tx{diag}\big(e^{+i\theta},e^{-i\theta},e^{+i\theta},e^{-i\theta}{}^{}\big)$ with $\theta$ being the scattering angle between $\bs{k}^{}_\pi$ and $\bs{k}^{}_a$ the three momenta of the pion and axion, respectively, $E^{}_\pi = \sqrt{\vkpi^2 +m_\pi^2}$ is the energy of the pion, $\Delta m = m^{}_\Delta - m^{}_N \simeq 293\,\tx{MeV}$ is the mass difference between $\Delta$ decuplet baryon and nucleon, and $\Gamma_\Delta \simeq 117\,\tx{MeV}$ is the decay width of the $\Delta(1232)$ baryon~\cite{ParticleDataGroup:2022pth}.\footnote{At finite temperatures, the decay width should depend on the background temperature~\cite{Ho:2015jva}.\,\,However, in our case $m^{}_\Delta \gg T$, the thermal effect on the decay width of the $\Delta$ baryon is very weak.\,\,For simplicity, we do not adopt the temperature-dependent decay width in our numerical calculations.}\,\,Using the following cross section formula in the laboratory frame, where an incident charged pion collides with a proton at rest
\begin{eqnarray}
\hs{-0.4cm}
\sigma^{}_{\pi^- p \to n a}
=
\mathop{\mathlarger{\int}} \hs{-0.15cm}
\frac{\dd^3 \bs{k}^{}_a}{(2\pi)^3 2E^{}_a}
\frac{\dd^3 \bs{k}^{}_n}{(2\pi)^3 2E^{}_n}
{}^{}(2\pi)^4 \delta^{(4)}\big(k^{}_\pi + k^{}_p - k^{}_a - k^{}_n\big)
\frac{\OL{\big|{\cal M}_{\pi^- p \to n a}\big|^2}}
{4\big[(k^{}_\pi \cdot k^{}_p{}^{})^2-({}^{}m^{}_\pi m^{}_N)^2\big]\raisebox{1pt}{$\hs{-0.05cm}^{1/2}$}}
\end{eqnarray}
with $k^{}_j = (E^{}_j, \bs{k}^{}_j)$ is the four-momenta of particle species $j$, the resultant cross section of $\pi^- + p \to n + a$ calculated in the HBChPT at large $m^{}_N$ expansion is then\footnote{In the HBChPT, the large $m^{}_N$ means that the nucleon mass is much bigger than the momentum and energy of pions, $m^{}_N \gg \vkpi,E^{}_\pi$.\,\,Thus, one can expand  physical observables in terms of $\vkpi/m^{}_N$ and $E^{}_\pi/m^{}_N$.}
\begin{eqnarray}
\sigma^{}_{\pi^- p \to n a}
\Eq
\frac{\Epi {}^{} m_N^2}{16{}^{}\pi f_\pi^2 f_a^2 {}^{} \vkpi} {}^{}{}^{}
{\cal G}^{}_{\hs{-0.02cm}a}(\vkpi) 
~,
\end{eqnarray}
where ${\cal G}^{}_{\hs{-0.02cm}a}(\vkpi)$ is a dimensionless quantity expressed by
\begin{eqnarray}\label{Ga}
{\cal G}^{}_{\hs{-0.02cm}a}(\vkpi)
\Eq
\frac{2 {}^{} g_A^2\big(2 {}^{} C_{+}^2 + C_{-}^2\big)}{3}
\bigg(\frac{\vkpi}{m^{}_N}\bigg)^{\hs{-0.13cm}2}
+
\CapiN^2\bigg(\frac{\Epi}{m^{}_N}\bigg)^{\hs{-0.13cm}2}
+
\frac{8\sqrt{2} {}^{}{}^{} g^{}_A \CapiN^{} {}^{} C^{}_-}{3}
\bigg(\frac{\vkpi}{m^{}_N}\bigg)^{\hs{-0.13cm}2}
\bigg(\frac{\Epi}{m^{}_N}\bigg)
\nn[0.1cm]
&&+{}^{}{}^{}
\frac{4{}^{}\CaND^2 {\cal C}^2}{81}
\frac{\Epi^2 \big(\Delta m^2 + 2 \Epi^2 + \bar{\Gamma}_\Delta^2\big)}
{\sx{1.0}{\big[} 
\big(\Delta m - \Epi^{} \big)\raisebox{0.5pt}{$^{\hs{-0.03cm}2}$} + \bar{\Gamma}_\Delta^2{}^{}{}^{}
\sx{1.0}{\big]}
{\sx{1.0}{\big[} 
\big(\Delta m + \Epi^{} \big)\raisebox{0.5pt}{$^{\hs{-0.03cm}2}$} + \bar{\Gamma}_\Delta^2{}^{}{}^{}
\sx{1.0}{\big]}}}
\bigg(\frac{\vkpi}{m^{}_N}\bigg)^{\hs{-0.13cm}2}
\nn[0.1cm]
&&-{}^{}{}^{}
\frac{8\sqrt{3} {}^{}{}^{} g^{}_A \CaND^{} {\cal C}}{27}
\frac{\Epi 
\sx{1.0}{\big[} 
\big(\Delta m^2 - \Epi^2 \big) \big(C^{}_+ \Delta m + C^{}_- \Epi \big) 
+ 
\bar{\Gamma}^2_\Delta \big(C^{}_+ \Delta m - C^{}_- \Epi \big) 
\sx{1.0}{\big]}}
{\sx{1.0}{\big[} 
\big(\Delta m - \Epi^{} \big)\raisebox{0.5pt}{$^{\hs{-0.03cm}2}$} + \bar{\Gamma}_\Delta^2{}^{}{}^{}
\sx{1.0}{\big]}
{\sx{1.0}{\big[} 
\big(\Delta m + \Epi^{} \big)\raisebox{0.5pt}{$^{\hs{-0.03cm}2}$} + \bar{\Gamma}_\Delta^2{}^{}{}^{}
\sx{1.0}{\big]}}}
\bigg(\frac{\vkpi}{m^{}_N}\bigg)^{\hs{-0.13cm}2}
\nn[0.1cm]
&&-{}^{}{}^{}
\frac{16\sqrt{6} {}^{}{}^{} \CapiN^{} {}^{} \CaND^{} {\cal C}}{27}
\frac{\Epi^2 \big(\Delta m^2 - \Epi^2 - \bar{\Gamma}^2_\Delta\big)}
{\sx{1.0}{\big[} 
\big(\Delta m - \Epi^{} \big)\raisebox{1pt}{$^{\hs{-0.03cm}2}$} + \bar{\Gamma}_\Delta^2{}^{}{}^{}
\sx{1.0}{\big]}
{\sx{1.0}{\big[} 
\big(\Delta m + \Epi^{} \big)\raisebox{1pt}{$^{\hs{-0.03cm}2}$} + \bar{\Gamma}_\Delta^2{}^{}{}^{}
\sx{1.0}{\big]}}}
\bigg(\frac{\vkpi}{m^{}_N}\bigg)^{\hs{-0.13cm}2}
\bigg(\frac{\Epi^{}}{m^{}_N}\bigg)
\end{eqnarray}
with $C^{}_{\pm} \equiv \big(C^{}_{ap} \pm C^{}_{an}\big)/2$, and $\bar{\Gamma}_\Delta = \Gamma_\Delta/2$.\footnote{To make our calculation result more reliable, we have also checked that $\,{\cal G}^{}_a$ at leading order in $1/m^{}_N$ using the Rarita-Schwinger propagator~\cite{Haberzettl:1998rw} is consistent with the decuplet propagator in the HBChPT~\cite{Jen:1991}.}\,\,Notice that the first, second, and fourth terms in Eq.\,\eqref{Ga} come from the nucleon-mediated, contact, and $\Delta$-mediated diagrams in Fig.\,\ref{fig:pipan}, respectively, and the other terms are the interference terms of those contributions.\,\,Further, the third term (last term) which is the interference term of the contact and nucleon-mediated ($\Delta$-mediated) diagrams is the subleading term ($\sim 1/m_N^3$) in Eq.\,\eqref{Ga} at large $m^{}_N$ expansion.\footnote{Note that our subleading terms in Eq.\,\eqref{Ga} are different from those in \cite{Choi:2021ign}.\,\,This is because they use the gamma matrix formalism in the relativistic quantum field theory, while we adopt the spin operator formalism in the HBChPT for the Lagrangian to compute the supernova axion emissivity.}

\begin{figure}[t!]
\centering
\hs{0.0cm}
\includegraphics[width=0.485\textwidth]{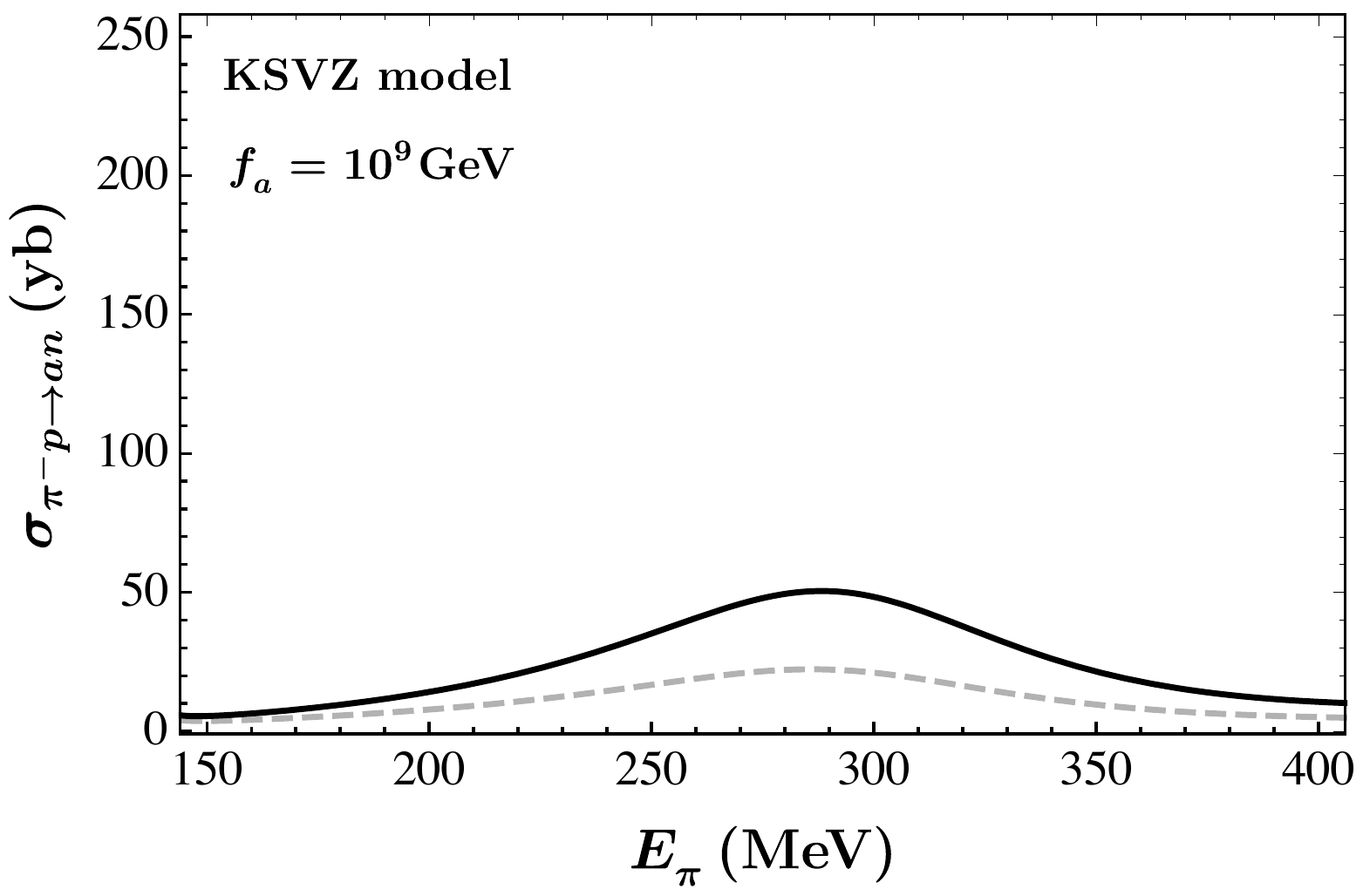}
\hs{0.0cm}
\includegraphics[width=0.485\textwidth]{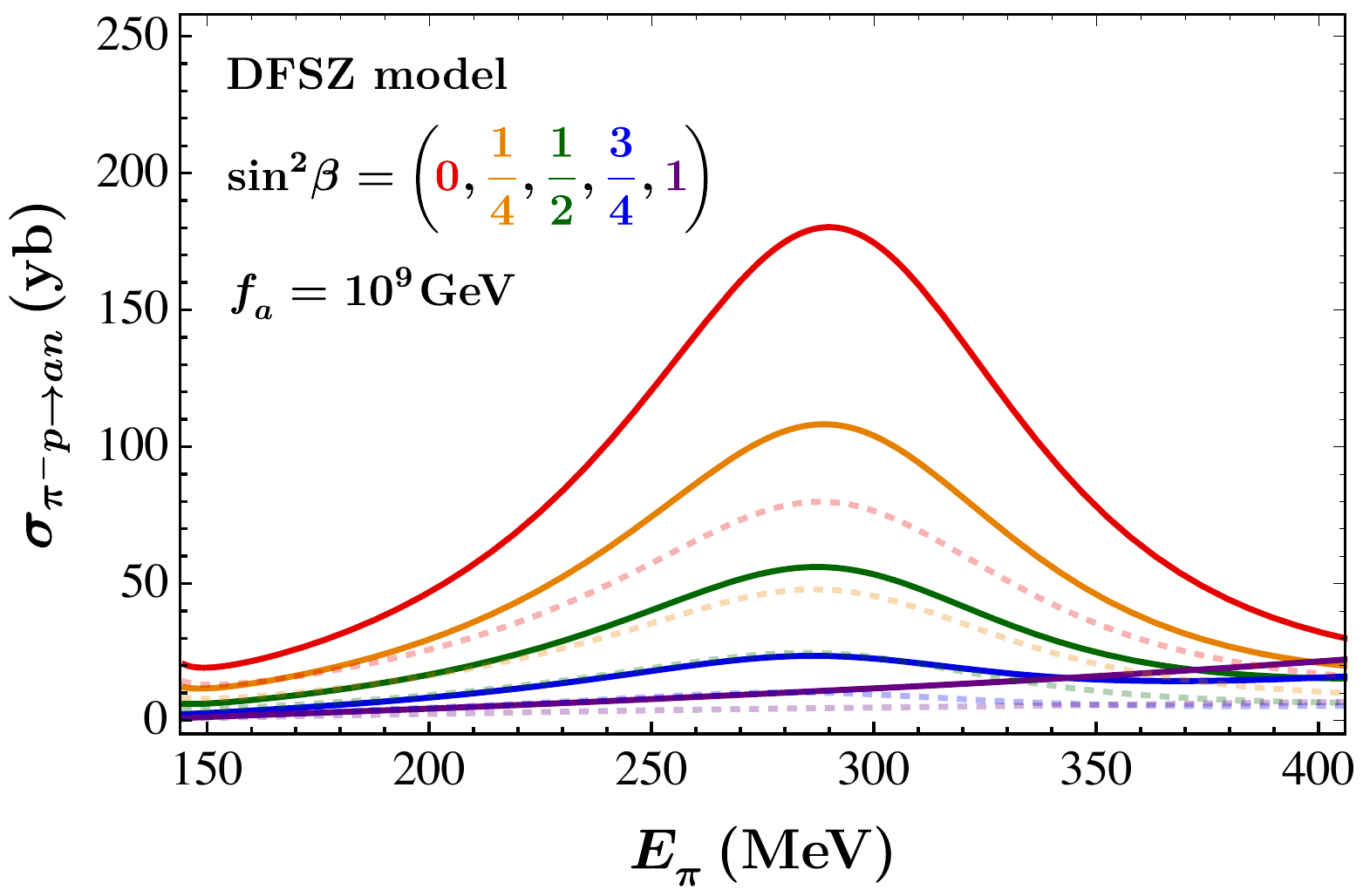}
\vs{-0.3cm}
\caption{Scattering cross section of $\pi^- + p \to a + n$ versus the energy of the incident pion in  the KSVZ model, and in the DFSZ model with different values of $\sin^2 \hs{-0.07cm} \beta$, where $\tx{yb} = 10^{-48}\,\tx{cm}^2$.\,\,In these figures, the solid lines indicate the scattering cross section in a large $m^{}_N$ limit, and the dashed lines represent the scattering cross section without a large $m^{}_N$ limit.\,\,Notice that the scattering cross section in the large $m^{}_N$ limit is bigger than the ones without a large $m^{}_N$ limit.}
\label{fig:csEpi}
\vs{0.0cm}
\end{figure}

We show in Fig.\,\ref{fig:csEpi} the scattering cross section of $\pi^- + p \to n + a$ as a function of $E^{}_\pi$ in the KSVZ and DFSZ models, where solid (dashed) curves are evaluated with (without) large $m^{}_N$ expansion.\,\,As anticipated, there is a resonance in the cross section when $E^{}_\pi \sim \Delta m$ and this is due to the $\Delta^0$-mediated diagram in Fig.\,\ref{fig:pipan}.\,\,In the case of the DFSZ model, one can see that the magnitude of the resonance becomes weaker as $\sin^2 \hs{-0.07cm} \beta \to 1$.\,\,This can be easily understood based on our calculation of the axion couplings to the decuplet baryons and nucleons in Eq.\,\eqref{CaNDn}, where $|\CaND(\sin^2 \hs{-0.07cm} \beta \to 1)| \sim 0.01$ which is suppressed compared with $|\CaND(\sin^2 \hs{-0.07cm} \beta \to 0)| \sim 1$.\,\,It is worth mentioning that the $\pi^- + p \to n + a$ cross section in the KSVZ model roughly corresponds to that in the DFSZ model with $\sin^2 \hs{-0.07cm} \beta \sim 1/2$ as can be observed in Fig.\,\ref{fig:csEpi}.\footnote{This correspondence of the KSVZ model and the DFSZ model when $\sin^2 \hs{-0.07cm} \beta = 1/2$ is also pointed out in~\cite{Vonk:2021sit}.}\,\,We expect that this correspondence will also occur in the supernova axion emissivity discussed in the next section.

\section{Supernova Axion Emission Rate with $\bs{\Delta(1232)}$ Resonance}\label{sec:5}

Given the axion-nucleon-$\Delta$ couplings derived in Sec.\,\ref{sec:3}, we can now evaluate the supernova axion emissivity of the process $\pi^- + p \to n + a$ with the contribution from the $\Delta$ resonance as shown in Fig.\,\ref{fig:csEpi}.\,\,The Feynman graphs of this axion emission process are the same as in Fig.\,\ref{fig:pipan}.

Following Ref.\,\cite{Choi:2021ign}, the supernova axion emission rate (the energy loss by axion radiations per unit volume and time) via the process $\pi ^- + p \to a + n$ is given by
\begin{eqnarray}
\dot{{\cal E}}_a
\Eq
\mathop{\mathlarger{\int}} \hs{-0.15cm}
\frac{\dd^3 \bs{k}^{}_\pi}{(2\pi)^3 2E^{}_\pi}
\frac{\dd^3 \bs{k}^{}_p}{(2\pi)^3 2E^{}_p}
\frac{\dd^3 \bs{k}^{}_a}{(2\pi)^3 2E^{}_a}
\frac{\dd^3 \bs{k}^{}_n}{(2\pi)^3 2E^{}_n}
{}^{}(2\pi)^4 \delta^{(4)}\big(k^{}_\pi + k^{}_p - k^{}_a - k^{}_n{}^{}\big)
\nn
&&\hs{0.4cm}\times
f^{}_\pi(|\bs{k}^{}_\pi|) f^{}_p(|\bs{k}^{}_p|)\big[1-f^{}_n(|\bs{k}^{}_n|)\big]
\big|{\cal M}^{}_{\pi^- p \to n a}{}^{}\big|^2 E^{}_a {}^{}
~,
\end{eqnarray}
where $f^{}_j(|\bs{k}^{}_j|) = 1/\big[e^{(E^{}_j-\mu^{}_j)/T)} \pm 1\big]$ is the Bose-Einstein $(-)$ or Fermi-Dirac $(+)$ distribution function with $\mu^{}_j$ being the chemical potential of particle species $j$, and $|{\cal M}_{\pi^- p \to n a}{}^{}|^2 = 2\,\OL{|{\cal M}_{\pi^- p \to n a}{}^{}|^2}$ is the squared matrix element summing over the initial and final nucleon spins.\,\,In the $1/m^{}_N$ expansion, the supernova axion emissivity with the $\Delta$ resonance contribution is calculated as\footnote{Our resulting supernova axion emission rate at leading order in $1/m^{}_N$ agrees with Ref.\,\cite{Choi:2021ign} without the axion to $\Delta$ interactions and slightly disagrees with Ref.\,\cite{Carenza:2020cis} without the axion contact and axion to $\Delta$ interactions.}
\begin{eqnarray}
\hs{-0.2cm}
\dot{{\cal E}}_a
\,=\,
\frac{z^{}_\pi z^{}_p}{f^2_\pi f^2_a}\sqrt{\frac{m_N^7 {}^{} T^{11}}{128{}^{}\pi^{10}}}
\mathop{\mathlarger{\int}_{\hs{-0.03cm}0}^\infty} \hs{-0.05cm} \dd x^{}_p
\frac{x_p^2 {}^{}{}^{} e^{x_p^2}}{\big(e^{x_p^2} + z^{}_n{}^{}\big)\big(e^{x_p^2} + z^{}_p{}^{}\big)}
\mathop{\mathlarger{\int}_{\hs{-0.03cm}0}^\infty} \hs{-0.05cm} \dd x^{}_\pi
\frac{x_\pi^2 {}^{} \epsilon^{}_\pi {}^{}{}^{}
{\cal F}^{}_{\hs{-0.02cm}a}(x^{}_\pi)}
{e^{{}^{}\epsilon^{}_\pi-y^{}_\pi} - z^{}_\pi}
~,
\end{eqnarray}
where $z^{}_j \,=\, e^{({}^{}{}^{}\mu_j - m_j)/T}$ is the fugacity of particle species $j$, 
\begin{eqnarray}
x^{}_p \,=\, \frac{|\bs{k}^{}_p|}{\sqrt{2{}^{}m^{}_N T}}
~,\quad
x^{}_\pi \,=\, \frac{\vkpi^{}}{T}
~,\quad
\epsilon^{}_\pi \,=\, \frac{\Epi^{}}{T}
~,\quad
y^{}_\pi \,=\, \frac{m^{}_\pi}{T}
\end{eqnarray}
with $\epsilon_\pi^2 = x_\pi^2 + y_\pi^2{}^{}$, ${\cal F}^{}_{\hs{-0.02cm}a}(x^{}_\pi) = {\cal G}^{}_{\hs{-0.02cm}a}(x^{}_\pi) + \Delta{\cal G}^{}_{\hs{-0.02cm}a}(x^{}_\pi)$ with ${\cal G}^{}_{\hs{-0.02cm}a}(x^{}_\pi) = {\cal G}^{}_{\hs{-0.02cm}a}(\vkpi)$ given in Eq.\,\eqref{Ga} and
\begin{eqnarray}
\hs{-0.5cm}
\Delta{\cal G}^{}_{\hs{-0.02cm}a}(x^{}_\pi)
\Eq
\frac{\sqrt{2} {}^{}{}^{} g^{}_A \CapiN^{} {}^{} C^{}_-}{3}
\frac{\Epi^4 - 3\big(\Delta m^2 - \bar{\Gamma}_\Delta^2\big) \Epi^2 + 
2\big(\Delta m^2 + \bar{\Gamma}_\Delta^2\big)\raisebox{1pt}{$^{\hs{-0.03cm}2}$}}
{\sx{1.0}{\big[} 
\big(\Delta m - \Epi^{} \big)\raisebox{1pt}{$^{\hs{-0.03cm}2}$} + \bar{\Gamma}_\Delta^2{}^{}{}^{}
\sx{1.0}{\big]}
{\sx{1.0}{\big[} 
\big(\Delta m + \Epi^{} \big)\raisebox{1pt}{$^{\hs{-0.03cm}2}$} + \bar{\Gamma}_\Delta^2{}^{}{}^{}
\sx{1.0}{\big]}}}
\bigg(\frac{\vkpi}{m^{}_N}\bigg)^{\hs{-0.13cm}2}
\bigg(\frac{\Epi^{}}{m^{}_N}\bigg)
~.
\end{eqnarray}
Here we have made use of Eq.\,\eqref{CaND} to simplify the above expression.

\begin{figure}[t!]
\centering
\hs{0.0cm}
\includegraphics[width=0.48\textwidth]{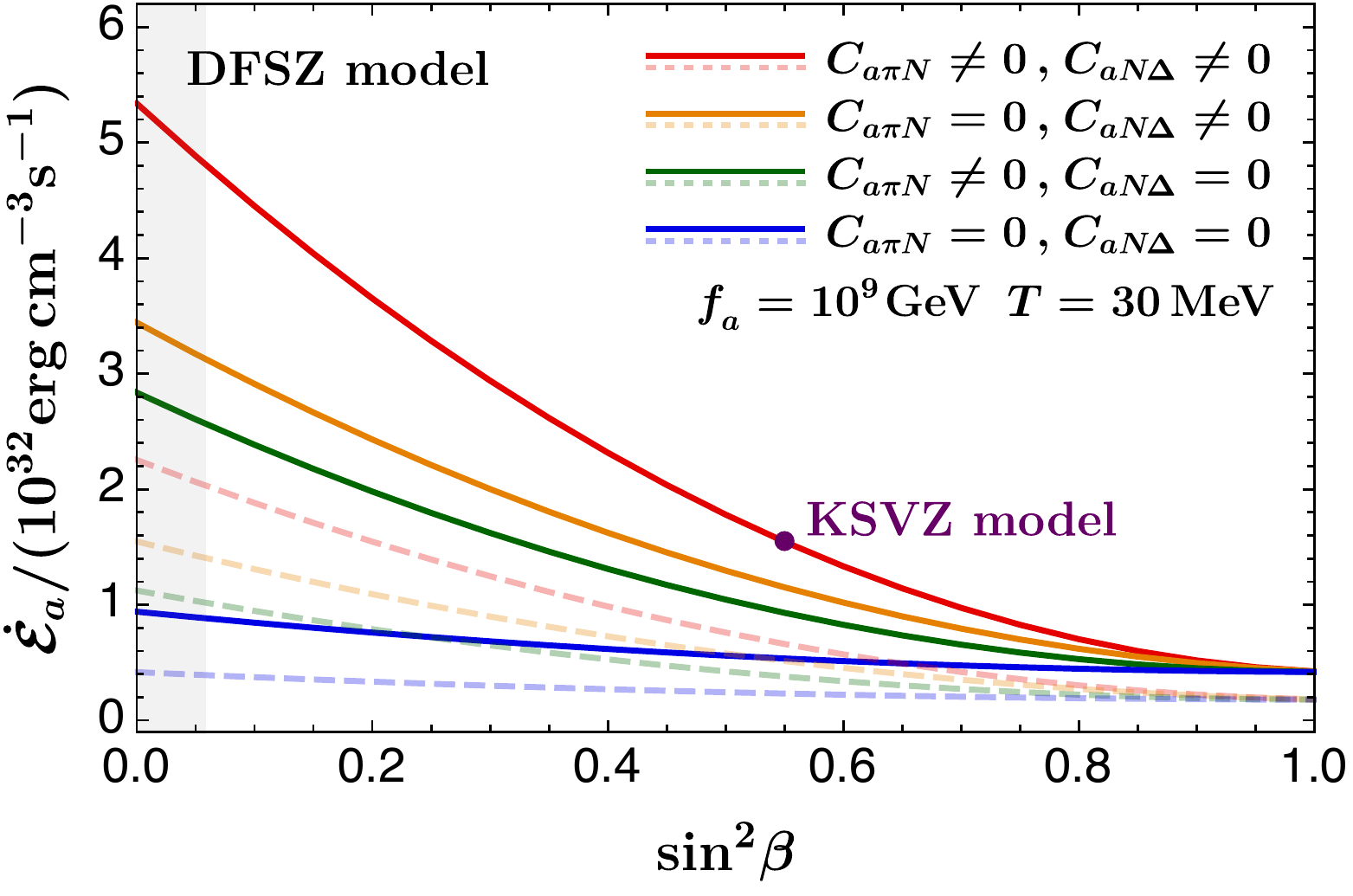}
\hs{0.1cm}
\includegraphics[width=0.489\textwidth]{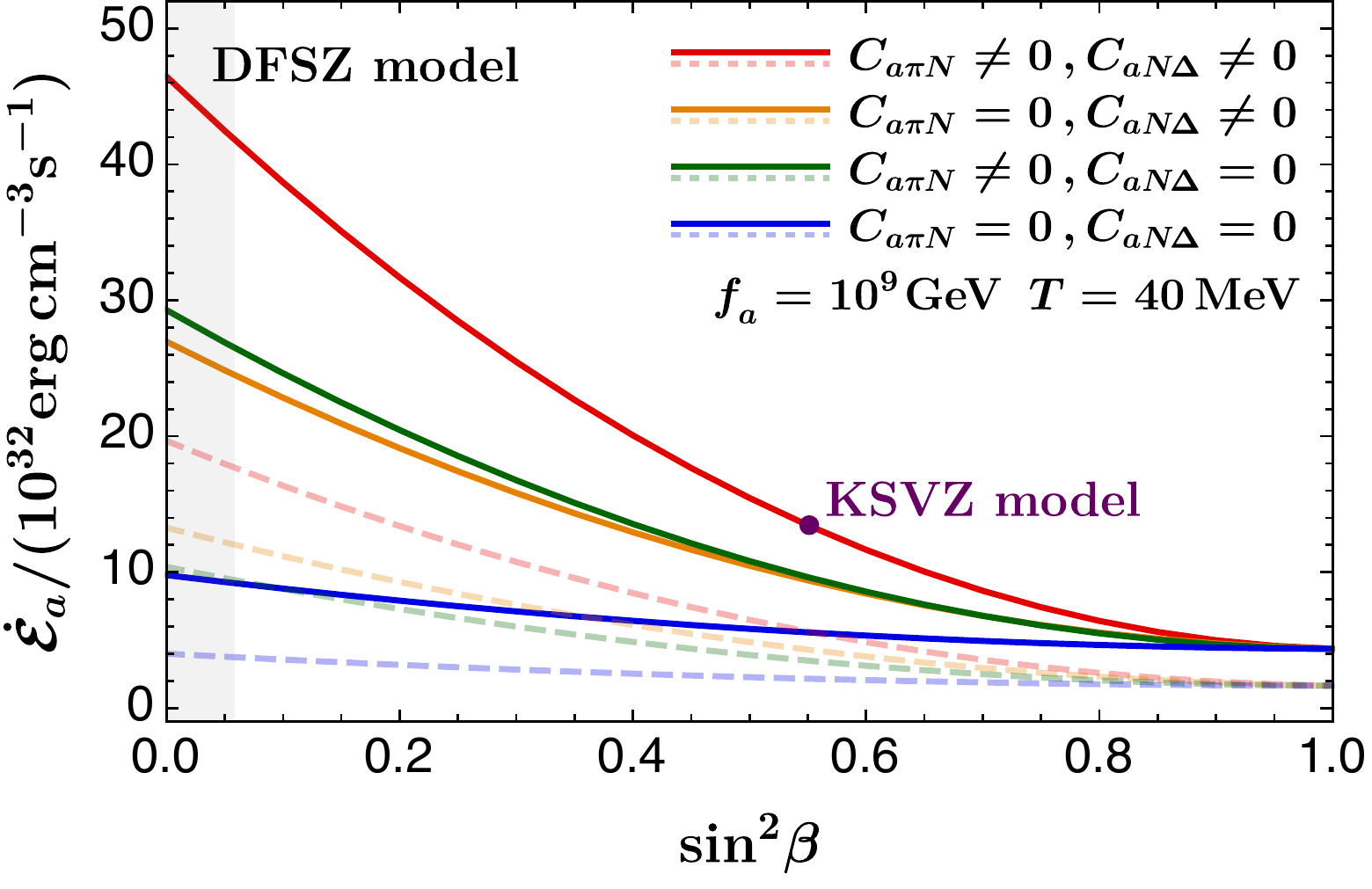}
\vs{-0.7cm}
\caption{Supernova axion emission rate versus $\sin^2 \hs{-0.07cm} \beta$ for two choices of the supernova temperature.\,\,To estimate the supernova axion emissivity, here we have used the fugacity values of the pion and nucleons in Ref.\,\cite{Fore:2019wib} at baryon density $n^{}_\tx{B} = 0.08\,\tx{fm}^{-3}$.\,\,Again, the solid lines indicate the supernova axion emission rate in a large $m^{}_N$ limit, and the dashed lines represent the supernova axion emission rate without a large $m^{}_N$ limit.}
\label{fig:easinb2}
\vs{0.0cm}
\end{figure}

We show in Fig.\,\ref{fig:easinb2} the supernova axion emission rate as a function of $\sin^2 \hs{-0.07cm} \beta$ for $T = 30\,\tx{MeV}$ and $40\,\tx{MeV}$ in the DFSZ model, where the distinction of the solid and dashed curves has been mentioned in the previous section.\,\,In these two figures, the gray band is excluded by tree-level unitarity of fermion scattering, where only $0.25 \lesssim \tan \beta \lesssim 170$ is allowed~\cite{Ferreira:2020bpb}.\,\,Notice that the upper bound of $\sin^2 \hs{-0.07cm} \beta$ is not evident in the figures since it is extremely close to 1.\,\,These bounds on $\tan \beta$ also prevent the SM quarks from being massless in the DFSZ model, where $m^{}_u \sim y^{}_u \upsilon \sin \beta$ and $m^{}_d \sim y^{}_d {}^{}{}^{} \upsilon \cos \beta$ with $\upsilon = (\upsilon_u^2 + \upsilon_d^2{}^{})^{1/2}$.\,\,In the case of the KSVZ model, the values of $\dot{{\cal E}}_a$ can be read from the curves with $\sin^2 \hs{-0.07cm} \beta \sim 1/2$  (the purple spots) of these figures, as pointed out in the previous section.\,\,By comparing these two figures, one can see that the contribution of the $\Delta$ resonance can be dominant over or comparable with that of the axion contact interaction for the typical supernova temperature.\,\,On the other hand, their contributions become negligible when $\sin^2 \hs{-0.07cm} \beta \to 1$ because $|C_{a\pi \hs{-0.02cm} N,{}^{}{}^{}aN \hs{-0.03cm} \Delta}(\sin^2 \hs{-0.07cm} \beta \to 1)| \ll |C_{a\pi \hs{-0.02cm} N,{}^{}{}^{}aN \hs{-0.03cm} \Delta}(\sin^2 \hs{-0.07cm} \beta \to 0)|$ according to Eqs.\,\eqref{CapiNn} and \eqref{CaNDn}.\,\,With the $\Delta$ resonance contribution, we see that the supernova axion emission rate can be enhanced at most by a factor of $\sim\,$2 for small $\sin^2 \hs{-0.07cm} \beta$ values compared with the earlier study in the presence of the axion-nucleon and axion-pion-nucleon contact interactions~\cite{Choi:2021ign}.

\begin{figure}[t!]
\centering
\vs{-0.35cm}
\includegraphics[width=0.48\textwidth]{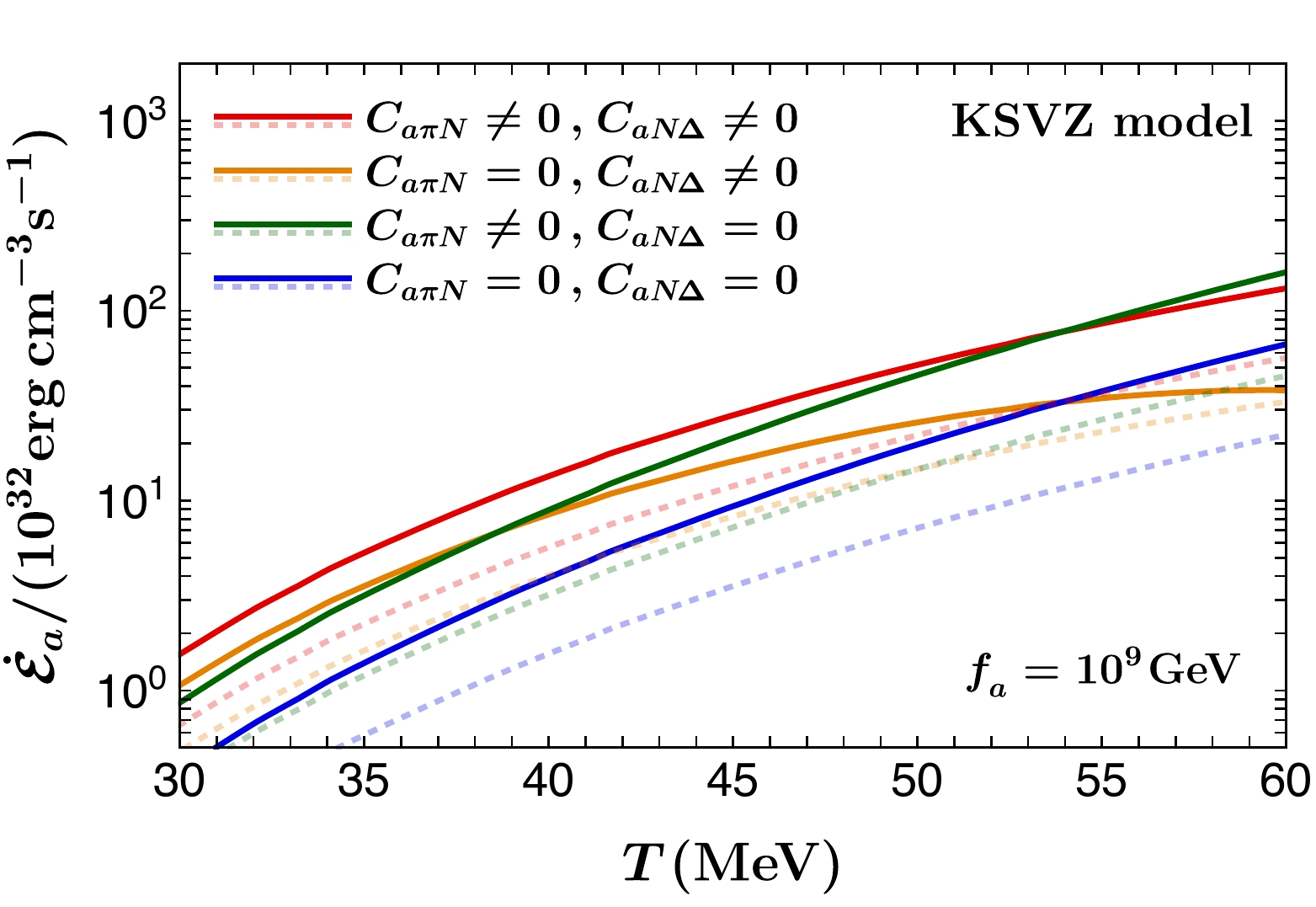}
\hs{0.2cm}
\includegraphics[width=0.48\textwidth]{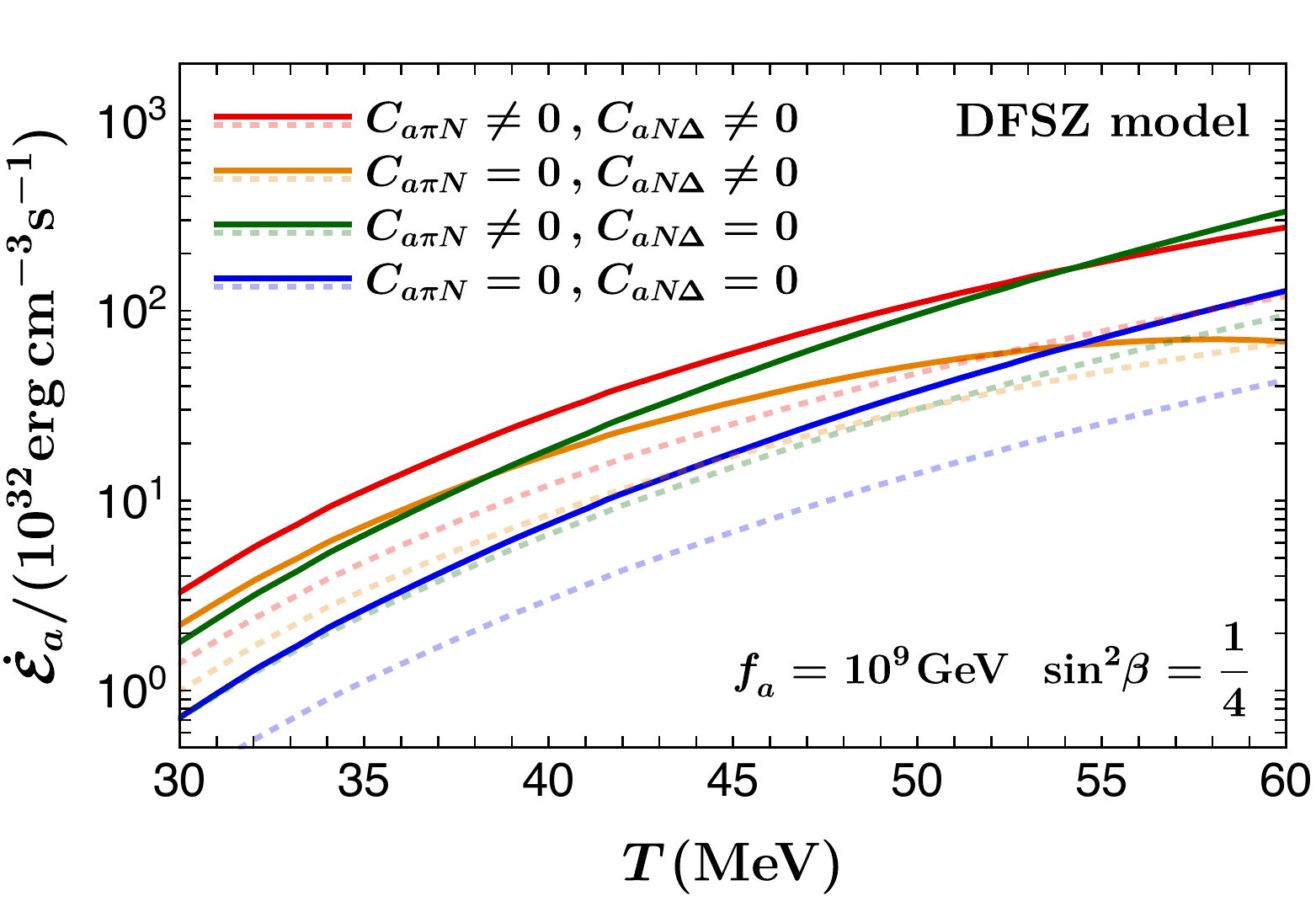}
\\[0.1cm]
\hs{0.01cm}
\includegraphics[width=0.48\textwidth]{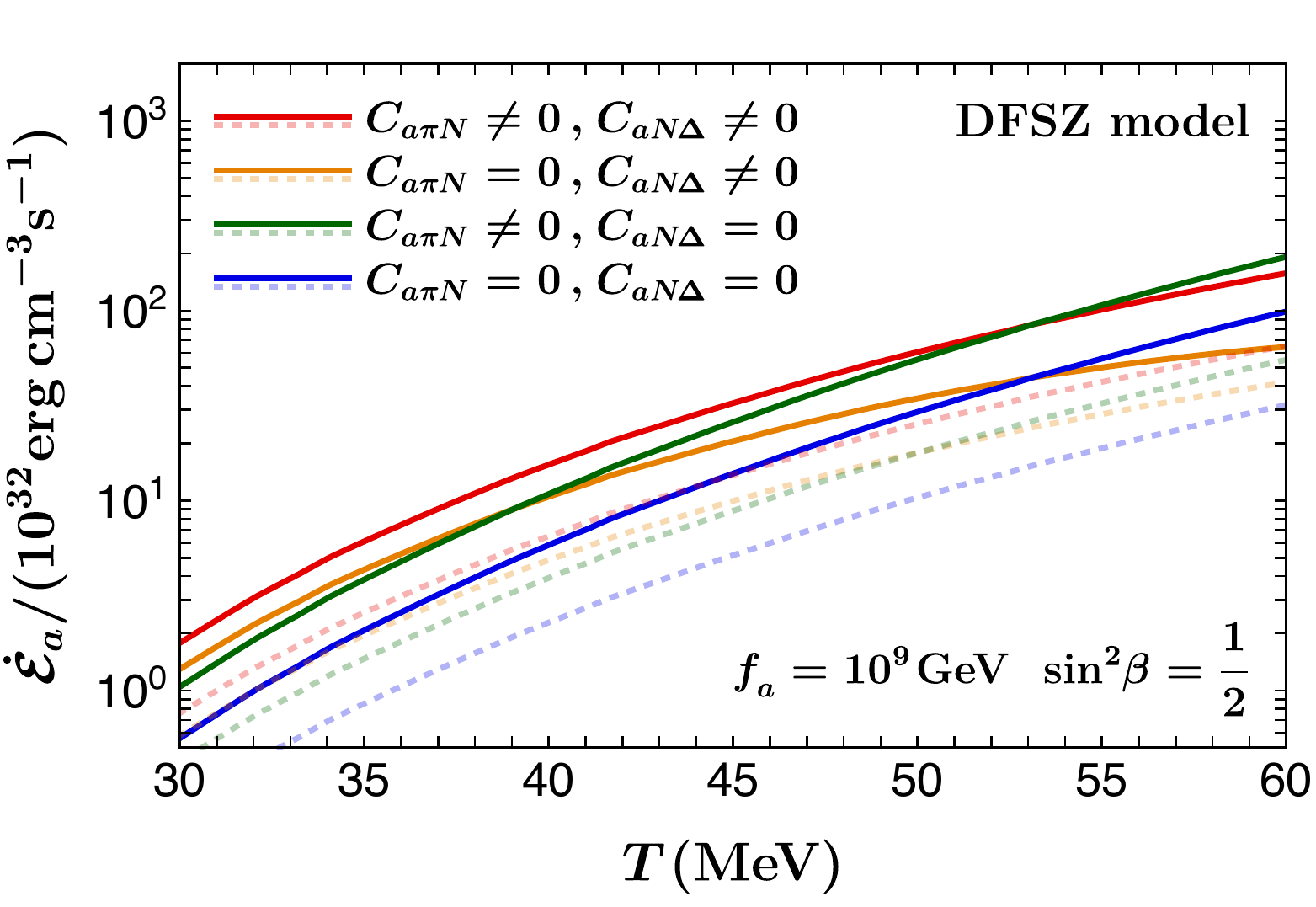}
\hs{0.2cm}
\includegraphics[width=0.48\textwidth]{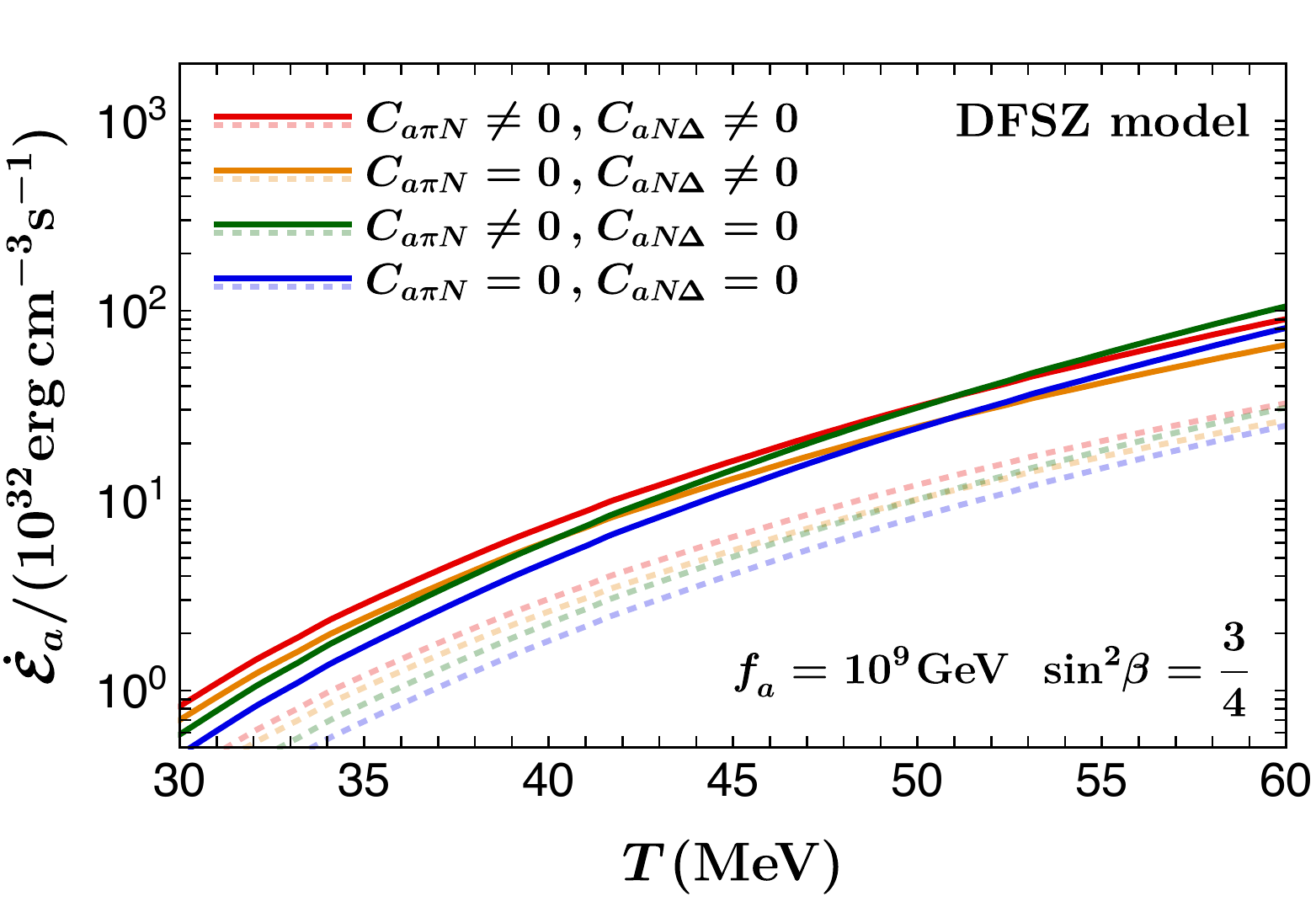}
\vs{-0.3cm}
\caption{Supernova axion emission rate as a function of $\,T$ in the KSVZ model, and in the DFSZ model with several choices of $\sin^2 \hs{-0.07cm} \beta$, where we have adopted the same fugacity values as in Fig.\,\ref{fig:easinb2}.\,\,The description of the solid and dashed lines is the same as in Figs.\,\ref{fig:csEpi} and \ref{fig:easinb2}.}
\label{fig:EavsT}
\vs{0.0cm}
\end{figure}

We also present in Fig.\,\ref{fig:EavsT} the supernova axion emission rate as a function of $\,T$ in the KSVZ model and the DFSZ model, where the choices of $\sin^2 \hs{-0.07cm} \beta$ in the DFSZ model satisfy the unitarity bounds.\,Again, the indication of the solid and dashed curves is the same as in Fig.\,\ref{fig:easinb2}.\,\,From these figures, we can see that the contribution of the $\Delta$ resonance is smaller (bigger) than that of the axion contact interaction if $\,T$ is higher (lower) than about $40\,\tx{MeV}$.\,\,Moreover, the $\Delta$ resonance contribution gives strongly destructive interference to the other contributions of the supernova axion emissivity at high supernova temperatures (${}^{}{}^{}T \gtrsim 55\,\tx{MeV}$).\,\,In the top left figure, one can notice that the supernova axion emission rate is enhanced by a factor of around 5\,(2) in the KSVZ model compared with the previous estimation including the axion-nucleon (and axion-pion-nucleon contact) interactions~\cite{Choi:2021ign}.\,\,Lastly, the enhancement of the supernova axion emission rate due to the $\Delta(1232)$ resonance contribution for the typical values of $\,T$ and small values of $\sin^2 \hs{-0.07cm} \beta$ in the DFSZ model has been mentioned in the previous paragraph.

\section{Discussion and Conclusions}\label{sec:6}

Before giving a conclusion of this work, let us comment on other new particle emission processes for supernovae.\,\,For instance, the dark photon emission from a supernova induced by the nucleon bremsstrahlung, $NN \to NN\gamma'$, can place the constraint on the kinetic mixing parameter and mass of the dark photon~\cite{Kazanas:2014mca}.\,\,Also, it has been shown in a recent paper~\cite{Shin:2022ulh} that the pion-induced Compton like process, $\pi^- + N \to N +\gamma'$, also plays a crucial role in the supernova dark photon emission due to the enhancement of the pion density inside supernovae.\,\,Therefore, we expect that the $\Delta(1232)$ resonance may also give a non-negligible contribution to this process as demonstrated in this work.\,\,We leave the estimation of the supernova dark photon emissivity induced by the Compton like process with the $\Delta(1232)$ resonance as a future investigation.

In this paper, we have estimated the energy loss rate from supernovae induced by the axion emission process as well as the axion production cross section including $\Delta(1232)$ resonance in the heavy baryon chiral perturbation theory.\,\,We have evaluated the supernova axion emissivity including axion-nucleon-$\Delta$ couplings which were neglected in the previous works.\,\,Since for the typical supernova temperatures, the energy of pion is $E^{}_\pi \sim 200\,\tx{MeV}$, the invariant mass of the $s$-channel mediator is somewhere in the middle of $\Delta(1232)$ and nucleon masses.\,\,Therefore, we cannot simply ignore the $\Delta(1232)$ baryon contributions to the supernova axion emission rate, as confirmed by explicit calculations demonstrated in this paper.\,\,We have also found that the supernova axion emission rate was overestimated by taking large $m^{}_N$ expansion in both DFSZ and KSVZ models.\,\,Thanks to the $\Delta(1232)$ resonance contribution, we have displayed that the supernova axion emissivity can be enhanced by a factor of 5\,(2) or so in the KSVZ model and up to a factor of about 4\,(2) in the DFSZ model 
for the small $\tan\beta$ values compared with the case only with the axion-nucleon 
(and axion-pion-nucleon contact) interactions.\,\,Finally, we notice that the $\Delta(1232)$ resonance can give a destructive contribution to the supernova axion emission rate at high supernova temperatures.

\section{Acknowledgments}\label{sec:7}

This work is supported by KIAS Individual Grants under Grants No.\,PG081201 (S.Y.H.), No.\,PG074202 (J.K.), and No.\,PG021403 (P.K.), and by National Research Foundation of Korea (NRF) Research Grant NRF-2019R1A2C3005009 (P.K., Jh.P.)

\end{document}